\documentclass[aps,pacs,twocolumn]{revtex4-1}
\usepackage[left=1.5cm,right=1.5cm,top=2.83cm,bottom=1cm]{geometry}
\usepackage{graphicx,epsfig}
\usepackage{amsmath}
\usepackage{hyperref}
\hypersetup{
     colorlinks   = true,
     citecolor    = magenta,
     linkcolor    = blue,
}

\begin{document}

\title{Percolation analysis of a disordered spinor Bose gas}
\author{Sk Noor Nabi}\email{sk.noor@iitg.ernet.in}
\author{Saurabh Basu} \email{saurabh@iitg.ernet.in}
\affiliation{Department of Physics, Indian Institute of Technology
  Guwahati, Guwahati, Assam 781039, India} \date{\today}

\begin{abstract}
We study the effects of an on-site disorder potential in a gas of
spinor (spin-1) ultracold atoms loaded in an optical lattice
corresponding to both ferromagnetic and antiferromagnetic spin
dependent interactions. Starting with a disordered spinor Bose-Hubbard
model (SBHM) on a two dimensional square lattice, we observe the
appearance of a Bose glass phase using the fraction of the lattice
sites having finite superfluid order parameter and non integer local
densities as an indicator. A precise distinction between three
different types of phases namely, superfluid (SF), Mott insulator (MI)
and Bose glass (BG) is done via a percolation analysis thereby
demonstrating that a reliable enumeration of phases is possible at
particular values of the parameters of the SBHM. Finally we present
the phase diagram based on the above information for both
antiferromagnetic and ferromagnetic interactions.
\end{abstract}

\keywords{latex-community, revtex4-1, aps, papers}

\maketitle

\section{Introduction}
\label{intro}
The physical realization of spinor Bose gases is severely impeded by
largely the instability of the internal atomic states. The excited
states typically decay through spontaneous emission on time scales
faster than that what is needed for the gas to equilibriate via
collisions. However, owing to the huge technological success, dilute
atomic gases with internal degrees of freedom have been
realized. Optically trapped spin-1 $^{23}Na$ is among the first ones
to be reported \cite{Stamper-Kurn,Stenger,Miesner,Tin,Wagner,Hou}
. The progress has been rapid since then with the discovery of spin-1
\cite{Wagner,Barrett} and spin-2 manifold of $^{87}Rb$
\cite{Hou}, \cite{Chang,Kuwamoto,Schmaljohann}.  A flurry of activities
have taken place since. It is beyond the scope of the present work to
furnish the large volume of literature that exists on the progress of
the study of spinor Bose gases. There is an extensive review by
Stamper-Kurn and Ueda on the subject \cite{Ueda}.  \\ \indent An
interesting offshoot of the study of phase diagram of spinor Bose
gases is to investigate the role of disorder present therein.  Since
inclusion of disorder in the context of an atomic gas loaded in an
optical lattice is experimentally possible via speckle or periodic
potentials \cite{Lye,White,Fallani,Damski,Niederle}, investigation of
its influence on the phase diagram is a pertinent task.  \\ \indent
Further, with the advent of techniques that can probe single atoms,
the accessibility of the quantum world has been revolutionized, which
can not be obtained by statistical sampling \cite{Bakr,Sherson}. This
has made possible imaging of the insulating phase of the interacting
atomic species of bosons on an optical lattice with a single atom and
single site resolution. Among other things, such studies will furnish
the population density of bosons (or the fluctuation therein) in
different phases and their local tunneling dynamics in an optical
lattice.  \\ \indent All of the above discussion would gain importance
in the context of studying a system of spinor bosons in an
inhomogeneous environment, that is in presence of disorder. The phase
diagram of the spinor particles is in general richer owing to the
different signs of the spin dependent interactions that render a polar
or a ferromagnetic nature for the superfluid phase
\cite{Tin}, \cite{Pai,Stefan}.  \\ \indent The
'$theorem\hspace{2mm}of\hspace{2mm}inclusions$' \cite{Pollet} forbids
a direct MI to a SF transition in presence of disorder. A BG phase
always intervenes in between.  Since a BG phase originates due to
arbitrarily large, although rarely occurring clusters of one type of
phase present amidst another, and as statistically rare events are
difficult to sample, elegant methods such as stochastic mean-field
theory \cite{Bissbort,Bissbort1,Neiderle1}, quantum Monte Carlo(QMC)
etc \cite{Neiderle1}, \cite{Lee,Makivic,Prokofev} were formulated and
used to study the BG phase.  \\ \indent In this work we employ a
single site mean field theory (SMFT) \cite{Sheshadri} which is very
extensively and most commonly used tool and apply it for studying a
Bose-Hubbard model (BHM) that recognizes the spin degree of freedom of
the atomic gas (we call it SBHM). Corresponding to the disorder free
scenario, the nature of phase transition in the antiferromagnetic and
ferromagnetic cases are ascertained via the average order parameter
and the local density fluctuation, that is the compressibility. The
nature of the phase transition for the odd (occupation density 1, 3,
5..) and the even (occupation density 2, 4, 6..)  Mott lobes are
different for the polar fluid, while this is untrue for the
ferromagnetic case ( see discussion on antiferromagnetic and
ferromagnetic cases in the next section).  \\ \indent In presence of
disorder, we define a simple physical quantity, $\chi$ defined as the
fraction of the sites with finite SF order parameter and non integer
occupation density that will be used as indicators for the MI, BG or
SF phases \cite{Apurba}.\\ \indent A percolation of the insulating
sites is seen to occur with patches of the superfluid order surviving,
thereby maintaining a zero macroscopic superfluid order, although
compressible. The percolation scenario is robustly tested and a
percolation threshold, implying the onset of a SF phase is obtained
with the help of Hoshen-Kopelman(HK) algorithm \cite{Hoshen}.
Equipped with all of the above, we map out a phase diagram for the
($F=$1) spinor bosons in a two dimensional square (optical) lattice,
where we specifically focus on the emergence and sustenance of the BG
phase in both antiferromagnetic and ferromagnetic cases.  \\ \indent
We organize our paper as follows. In the next section, we briefly
review the mean field theory for the SBHM in the inhomogeneous case
(in presence of disorder). In section III, we present the results for
the average SF order parameter and the compressibility in both the
antiferromagnetic and ferromagnetic cases. Further, we study the
variation of $\chi$ in presence of disorder and the finite size
scaling properties. We obtain the phase diagram based on this
information and present it thereafter. Finally in section IV we
outline our concluding remarks.
 
\section{Model}
\begin{figure}
  \centerline{
    \hfill \psfig{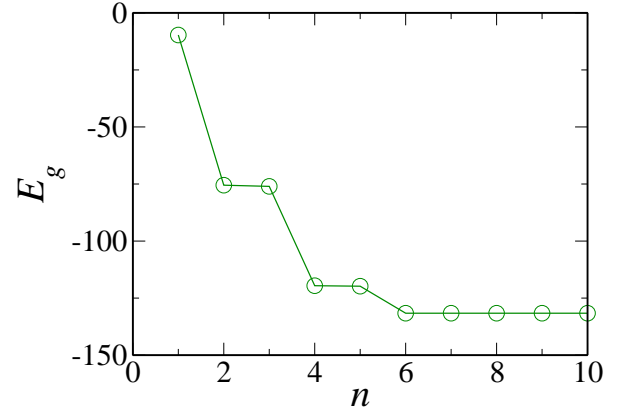}
    \hfill}
  \caption{The variation of the equilibrium ground state energy,
    $E_{g}$ with occupation number $n$.}
\label{eg}
\end{figure}

The behavior of ultracold atoms loaded in an optical lattice with
hyperfine spin $F$=1 can be well described by a SBHM \cite{Imambekov} as,
\begin{eqnarray}
\hat{H}=&-&t\sum\limits_{<i,j>}\sum\limits_{\sigma}(\hat{a}^{\dagger}_{i\sigma}\hat{a}_{i\sigma}+
h.c)+
\sum\limits_{i}(\mu-\epsilon_{i})\hat{n}_{i}\nonumber\\ &+&\sum\limits_{i}\frac{U_{0}}{2}\hat{n}_{i}(\hat{n}_{i}-1)
+\frac{U_{2}}{2}({\bf{S}}^{2}_{i}-2\hat{n}_{i})
\label{bhm1}
\end{eqnarray}
where $t$ is the tunneling matrix, $\langle i,j\rangle$ are the
nearest neighbour sites, $\mu$ is the chemical potential, $U_{0}$
describes the spin independent on-site interaction and $U_{2}$ is spin
dependent interaction which arises due to the difference in scattering
lengths $a_{0}$ and $a_{2}$ corresponding to $S$=0 and $S$=2 channels
and $U_{0}$, $U_{2}$ are related to the scattering length by,
$U_{0}=(4\pi \hbar^{2}/M)((a_{0}+2a_{2})/3)$ and $U_{2}$=
$(4\pi\hbar^{2}/M)((a_{2}-a_{0})/3)$. The spin dependent interaction,
$U_{2}$ can have either positive or negative signs depending upon
whether $a_{2}>a_{0}$ or $a_{2}<a_{0}$. Thus $U_{2}$ is
antiferromagnetic (where the SF phase is polar) for $a_{2}>a_{0}$ and
ferromagnetic (where the SF phase being ferromagnetic) for
$a_{2}<a_{0}$. The total spin at site $i$ is
${\bf{S}}_{i}=\hat{a}^{\dagger}_{i\sigma}
{\bf{F}}_{\sigma\sigma'}\hat{a}_{i\sigma'}$ where
${\bf{F}}_{\sigma\sigma'}$ are the components of spin-1 matrices and
$\sigma$=+1, 0, -1. The particle number operator,
$\hat{n}_{i}=\sum\nolimits_{\sigma}\hat{n}_{i\sigma}$,
$\hat{n}_{i\sigma}=\hat{a}_{i\sigma}^{\dag}\hat{a}_{i\sigma}$ where
$\hat{a}^{\dagger}_{i\sigma}(\hat{a}_{i\sigma})$ is the boson creation
(annihilation) operator at a site $i$. $\epsilon_{i}$ is the on-site
disorder is introduced by the parameter $i$, which is randomly chosen
from a box distribution extended over $[-\Delta,\Delta]$ where
$\Delta$ is the strength of the disorder and an important parameter of
our work. The components of the spin operator are given by,
\begin{eqnarray*}
\hat{S}_{iz}&=&\hat{a}^{\dagger}_{i+}\hat{a}_{i+}-\hat{a}^{\dagger}_{i-}\hat{a}_{i-}=\hat{n}_{i+}-\hat{n}_{i-}\\
\hat{S}_{ix}&=&\frac{1}{\sqrt{2}}(\hat{a}^{\dagger}_{i0}\hat{a}_{i+}+\hat{a}^{\dagger}_{i+}
\hat{a}_{i0}+\hat{a}^{\dagger}_{i0}\hat{a}_{i-}+\hat{a}^{\dagger}_{i-}\hat{a}_{i0})\\
\hat{S}_{iy}&=&\frac{i}{\sqrt{2}}(\hat{a}^{\dagger}_{i0}\hat{a}_{i+}-\hat{a}^{\dagger}_{i+}
\hat{a}_{i0}-\hat{a}^{\dagger}_{i0}\hat{a}_{i-}+\hat{a}^{\dagger}_{i-}\hat{a}_{i0})
\end{eqnarray*}
\begin{equation}
\begin{split}
S^{2}_{i}&=2\hat{n}_{i+}\hat{n}_{i0}+2\hat{n}_{i+}\hat{n}_{i0}+\hat{n}_{i+}+2\hat{n}_{i0}
+\hat{n}_{i-}+\hat{n}^{2}_{i+}\\
&-2\hat{n}_{i+}\hat{n}_{i-}+\hat{n}^{2}_{i-}+2\hat{a}^{\dagger}_{i+}\hat{a}^{\dagger}_{i-}
\hat{a}^{2}_{i0}+2\hat{a}_{i+}\hat{a}_{i-}\hat{a}^{\dagger2}_{i0}
\end{split}
\end{equation}
To solve Eq.(\ref{bhm1}), we shall employ SMFT that decouples the
system Hamiltonian into sum of the single site Hamiltonians. The
hopping part of the Hamiltonian can be decoupled as
\cite{Pai,Oosten},
\begin{eqnarray}
\hat{a}^{\dagger}_{i\sigma}\hat{a}_{j\sigma}=\langle
\hat{a}^{\dagger}_{i\sigma} \rangle \hat{a}_{j\sigma}
+\hat{a}^{\dagger}_{i\sigma}\langle
\hat{a}^{\dagger}_{j\sigma}\rangle-\langle
\hat{a}^{\dagger}_{i\sigma}\rangle \langle
\hat{a}^{\dagger}_{j\sigma}\rangle+\delta a^{\dagger}_{i\sigma}\delta
a_{j\sigma}
\label{deco}
\end{eqnarray}
where $\langle \hat{O}\rangle$ denotes the equilibrium value of an
operator, $\hat{O}$ and $\delta a_{j\sigma}= \hat{a}_{j\sigma}-\langle
\hat{a}_{j\sigma} \rangle$ denotes the fluctuations in
$\hat{a}_{i\sigma}$. Now introducing the superfluid order parameter at
site $i$ as,
\begin{equation}
\psi_{i\sigma}\equiv \langle
\hat{a}^{\dagger}_{i\sigma}\rangle=\sqrt{n_{s\sigma}}\xi^{*}_{\sigma}
\equiv \langle \hat{a}_{i\sigma}\rangle=\sqrt{n_{s\sigma}}\xi_{\sigma}
\end{equation}
where $\psi_{i}$=$\sqrt{\psi^{2}_{i\sigma}}$=
$\sqrt{\psi^{2}_{i+}+\psi^{2}_{i0}+\psi^{2}_{i-}}$ and $n_{s\sigma}$
is the superfluid density and $\xi_{\sigma}$ is a normalized spinor
obeying
$\sum\nolimits_{\sigma}\xi^{*}_{\sigma}\xi_{\sigma}=1$.\\ \indent It
is obvious that all spinors are related to each other by a gauge
transformation, $e^{i\theta}$ and spin rotations,
$u(\alpha,\beta,\tau)=e^{-iS_{z}\alpha}e^{-iS_{y}\beta}e^{-iS_{z}\tau}$
where $(\alpha,\beta,\tau)$ are the Euler angles. The behavior of the
SF order parameter is different in the antiferromagnetic and ferromagnetic
regimes and are enumerated in the following
\cite{Ho,Law,Koashi}. \\ (i) Antiferromagnetic case: when $U_{2}>0$,
the spinor, $\xi_{\sigma}$ is given by,
\begin{equation}
\xi=e^{i\theta}u
\begin{pmatrix}
  0  \\
  1  \\
  0 
 \end{pmatrix}=e^{i\theta}
\begin{pmatrix}
 -\frac{1}{\sqrt{2}}e^{-i\alpha} sin \beta  \\
   cos\beta \\
 \frac{1}{\sqrt{2}}e^{i\alpha} sin \beta
 \end{pmatrix}
\end{equation}
\\If $\alpha$=$\beta$=$\frac{\pi}{2}$ then $\psi_{+}$=$\psi_{-}\ne 0$,
$\psi_{0}=0$ or $\alpha$=$[0,2\pi]$ and $\beta$=0 or $\pi$ then
$\psi_{+}$=$\psi_{-}$=0, $\psi_{0}\ne 0$. \\(ii) Ferromagnetic case:
when $U_{2}<0$, $\xi_{\sigma}$ is given by,
\begin{equation}
\xi=e^{i\theta}u
\begin{pmatrix}
  1  \\
  0  \\
  0 
 \end{pmatrix}=e^{i\theta}
\begin{pmatrix}
e^{-i\alpha} cos^2 \beta/2  \\
\sqrt{2}\hspace{2mm}cos\beta/2 \hspace{2mm}sin\beta/2 \\
e^{i\alpha} sin^2 \beta/2
 \end{pmatrix}
\end{equation}
\\If $\beta=\pi/2$ and $\alpha=0$ then $\psi_{+}=\psi_{-}$,
$\psi_{0}=\sqrt{2}\psi_{+}$. \\ \indent Now neglecting the quadratic
fluctuations and substituting the superfluid order parameter in
Eq.(\ref{bhm1}), the SBHM can be written as a sum of single site
Hamiltonians as,
\begin{equation*}
H=\sum\nolimits_{i}H^{MF}_{i} 
\end{equation*}
where
\begin{equation}
\begin{split}
H^{MF}_{i}&=-t\sum\limits_{\sigma}(\phi^{*}_{i\sigma}\hat{a}_{i\sigma}+\phi_{i\sigma}
\hat{a}^{\dagger}_{i\sigma})+t\sum\limits_{\sigma}\phi^{*}_{i\sigma}\psi_{i\sigma}
\\ &-(\mu-\epsilon_{i})\hat{n}_{i}+\frac{U_{0}}{2}\hat{n}_{i}(\hat{n}_{i}-1)+\frac{U_{2}}{2}({\bf{S}}^{2}_{i}-2\hat{n}_{i})
\label{mf}
\end{split}
\end{equation}
where $\phi_{i\sigma}=\sum\nolimits_{j}\psi_{j\sigma}$. The sum $j$
includes all nearest neighbors at the site $i$ of a square lattice with
$z=4$, $z$ being the coordination number.  \\ \indent Our main task is
to find the SF order parameter, $\psi_{i\sigma}$ by diagonalizing the
mean field Hamiltonian $H^{MF}_{i}$ in Eq.(\ref{mf}) in the site
occupation number basis, $|\hat{n}_{i\sigma}\rangle$. For that, we
first have to find the matrix elements of $H^{MF}_{i}$, which is done
via,
\begin{equation}
\langle
\hat{n}_{i+},\hat{n}_{i0},\hat{n}_{i-}|H^{MF}_{i}|\hat{n}'_{i+},\hat{n}'_{i0},\hat{n}'_{i}\rangle=h^{d}_{i}+
h^{od}_{i}
\label{mf1}
\end{equation}
where the $h^{d}_{i}$ part includes the matrix elements coming
from the diagonal part of the mean field Hamiltonian and reads as,
\begin{equation}
\begin{split}
h^{d}_{i}&=\Big[t\sum\nolimits_{\sigma}\phi^{*}_{i\sigma}\psi_{i\sigma}-(\mu-\epsilon_{i})(n_{i+}+n_{i0}+n_{i-})
\\ &+\frac{U_{0}}{2}(n_{i+}+n_{i0}+n_{i-})(n_{i+}+n_{i0}+n_{i-}-1)\\ &+\frac{U_{2}}{2}\big(2n_{i+}n_{i0}+2n_{i+}n_{i0}+n_{i+}+2n_{i0}+n_{i-}+n^{2}_{i+}
\\ &-2(n_{i+}+n_{i+}+n_{i+})\big)\Big]\delta_{n_{i+},n'_{i+}}\delta_{n_{i0},n'_{i0}}\delta_{n_{i-},n'_{i-}}
\end{split}
\end{equation} 
While the off diagonal part $h^{od}_{i}$ is,
\begin{equation}
\begin{split}
h^{od}_{i}&=t\Big[\phi^{*}_{i+}\sqrt{n_{i+}}\delta_{n_{i+},n'_{i+}+1}\delta_{n_{i0},n'_{i0}}\delta_{n_{i-},n'_{i-}}
\\ &+\phi^{*}_{i0}\sqrt{n_{i0}}\delta_{n_{i+},n'_{i+}}\delta_{n_{i0},n'_{i0}-1}\delta_{n_{i-},n'_{i-}}
\\ &+\phi^{*}_{i-}\sqrt{n_{i-}}\delta_{n_{i+},n'_{i+}}\delta_{n_{i0},n'_{i0}}\delta_{n_{i-},n'_{i-}-1}+h.c\Big]
\\ &+\frac{U_{2}}{2}\Big[\sqrt{n_{i0}(n_{i0}-1)(n_{i+}+1)(n_{i-}+1)}\\ &\delta_{n_{i+},n'_{i+}+1}\delta_{n_{i0},n'_{i0}-2}\delta_{n_{i-},n'_{i-}+1}+h.c\Big]
\end{split}
\end{equation}
\indent Hence we diagonalize the matrix in Eq.(\ref{mf1}) to obtain
the ground state energy $E^{i}_{g}(\psi_{i\sigma})$ and the
eigenfunction $\Psi_{g}(\psi_{i\sigma})$ starting with some some guess
value for $\psi_{i\sigma}$. Now from the updated wave function
$\Psi_{g}(\psi_{i\sigma})$, we compute the SF order parameter using,
\begin{equation} 
\psi_{i\sigma}=\langle\Psi_{g}(\psi_{i\sigma})|\hat{a}_{i\sigma}|\Psi_{g}(\psi_{i\sigma})\rangle
\end{equation} 
\\Using this new value of $\psi_{i\sigma}$, we again reconstruct the
matrix and repeat the diagonalization procedure until self consistency
condition is reached \cite{Sheshadri}. \\ \indent The variation of the
equilibrium ground state energy, $E_{g}$ with occupation number, $n$
shows that it almost stabilizes corresponding to $n$=7, for which
$\langle$$\hat{n}_{i+},\hat{n}_{i0},\hat{n}_{i-}|H^{MF}_{i}|\hat{n}'_{i+},\hat{n}'_{i0},
\hat{n}'_{i}$$\rangle$ is a $120\times120$ matrix [see Fig.\ref{eg}]. The
dimension is obtained by considering all possible combination of
$\hat{n}_{i+}$, $\hat{n}_{i0}$, $\hat{n}_{i-}$ such that,
$\hat{n}_{i+}+\hat{n}_{i0}+\hat{n}_{i-}$=$\hat{n}_{i}$. The local
density, $\rho_{i}$ and compressibility, $\kappa_{i}$ can also be
computed using,
\begin{equation}
\rho_{i}=\langle\Psi_{g}(\psi_{i\sigma})|\hat{n}_{i\sigma}|\Psi_{g}(\psi_{i\sigma})\rangle;\hspace{2mm}
 \kappa_{i}=\langle \rho^{2}_{i} \rangle-\langle \rho_{i} \rangle^{2}
\end{equation}
\\ \indent In the clean state, that is when $\Delta/U_{0}$=0, the mean
field Hamiltonian is homogeneous and hence the SF order parameter is
uniform throughout the entire lattice. In the strong interaction limit
($U_{0}\gg t$), the SF order parameter vanishes indicating that the
system is in the MI regime where the atoms are forbidden from
tunneling into neighbouring sites and thereby are localized to
individual sites with a fixed number of bosons per site. Thus the MI
phase is characterized by zero SF order parameter and zero
compressibility with a finite gap in the particle hole excitation
spectrum. \\ \indent In the weak interaction limit ($t \gg U_{0}$),
the SF order parameter shows finite value whence the atoms can easily
tunnel between the nearest neighbour sites. In the superfluid phase,
atoms are delocalized over the entire lattice and the ground state is
a coherent state, which in experiments is seen to form a constructive
interference pattern in the momentum space. Thus the superfluid phase
is characterized by non zero order parameter, non integer boson
density and finite compressibility with no particle hole excitation
gap in the spectrum.  \\ \indent In presence of disorder, the mean
field Hamiltonian becomes inhomogeneous and hence the SF order
parameter will vary from one site to another. Further, in the pure
case, the system can make direct phase transition from MI to SF phase
controlled by the system parameter $t$ and $U_{0}$. But as soon as
disorder is present, the BG phase always intervenes between the SF and
MI phases. The BG phase is defined as the zero SF order parameter with
no gap in the particle hole excitation spectra. Thus the three types
of phases can be characterized as, \\(i) SF phase: $\psi_{i} \ne 0$,
$\rho_{i}\ne $integer and $\kappa_{i}>0$ \\(ii) BG phase:
$\psi_{i}=0$, $\rho_{i}\ne$ integer and $\kappa_{i}>0$ \\(iii) MI
phase: $\psi_{i}=0$, $\rho_{i}=$integer and $\kappa_{i}=0$ \\ \indent
In order to see the effects of disorder, we first solve our mean field
Hamiltonian, Eq.(\ref{mf}) self consistently on a two dimensional (2D)
square lattice of size $L\times L$ where we have considered $L=30$
throughout our calculations (unless mentioned otherwise) to obtain the
averaged value of $\bar\psi$ and $\bar\rho$ as,
\begin{equation}
\bar\psi=\Big[\frac{1}{L^{2}}\sum\limits_{i=1}^{L^{2}}\psi_{i}\Big]_{sample} \hspace{2mm} and \hspace{2mm} 
\bar\kappa=\Big[\frac{1}{L^{2}}\sum\limits_{i=1}^{L^{2}}\kappa_{i}\Big]_{sample}
\end{equation}
where sample in the subscript refer to the fact that results are
averaged over different realizations of disorder. Here we have used a
maximum of 40 different disorder realizations and confirmed that the
fluctuations are satisfactorily equilibriated.
\section{Results}
\label{rslt}
\indent In this section we shall present our results by numerically
solving the mean field Hamiltonian, Eq.(\ref{mf1}) in the $zt-\mu$
plane while keeping $U_{0}$=1 and considered different values of spin
dependent interaction $U_{2}$.  Since
$U_{2}/U_{0}$=$(a_{2}-a_{0})/(a_{0}+2a_{2})$, so $U_{2}/U_{0}$ extends
from -1 when $a_{2}$=0 to 0.5 when $a_{0}$=0. In our work, we choose
$U_{2}/U_{0}$=0.1 (less than the maximum value 0.5) to study the odd
and even MI lobes. It may be noted that at $U_{2}/U_{0}\ge 0.5$, the
odd MI lobes are completely consumed by the even MI lobes. For the
ferromagnetic case, any value of $U_{2}$ in the interval [-1, 0] is
seen to yield qualitatively similar results and we choose
$U_{2}/U_{0}$=-0.2.  \\ \indent The other important parameter is the
disorder strength, $\Delta/U_{0}$. Since the MI phase is characterized
by finite energy gap in their energy spectrum, in the atomic limit
($t$=0) the width for the MI lobe is energy gap between the upper and
lower values of the chemical potential, $\mu$ corresponding to the
particle and hole excitations \cite{Fisher}. In the antiferromagntic
case, the width of the odd MI lobes is
$E_{g}$=$\mu_{+}-\mu_{-}$=$U_{0}-2U_{2}$, while for the even MI lobes,
it is $E_{g}$=$U_{0}+2U_{2}$ \cite{Lacki}. In order to go from a
gapped MI phase to a gappless BG phase, the disorder strength,
$\Delta/U_{0}$ should be greater or equal to the width of the
respective MI lobes. Since we are using a box disorder from -$\Delta$
to $\Delta$, so the critical disorder strength, $\Delta_{c}$ should
extend from -$E_{g}/2$ to +$E_{g}/2$ that is $\Delta_{c}/U_{0}$=0.4
for the odd MI lobes and $\Delta_{c}/U_{0}$=0.6 for the even MI lobes
corresponding to $U_{2}/U_{0}$=0.1.  Similarly for the ferromagnetic
case, the energy gap corresponds to $E_{g}$=$U_{0}+U_{2}$ and hence
the critical value is $\Delta_{c}/U_{0}$=0.4 corresponding to
$U_{2}/U_{0}$=-0.2. Using this parameter values, we shall present our
results systematically in the following.
\begin{figure*}
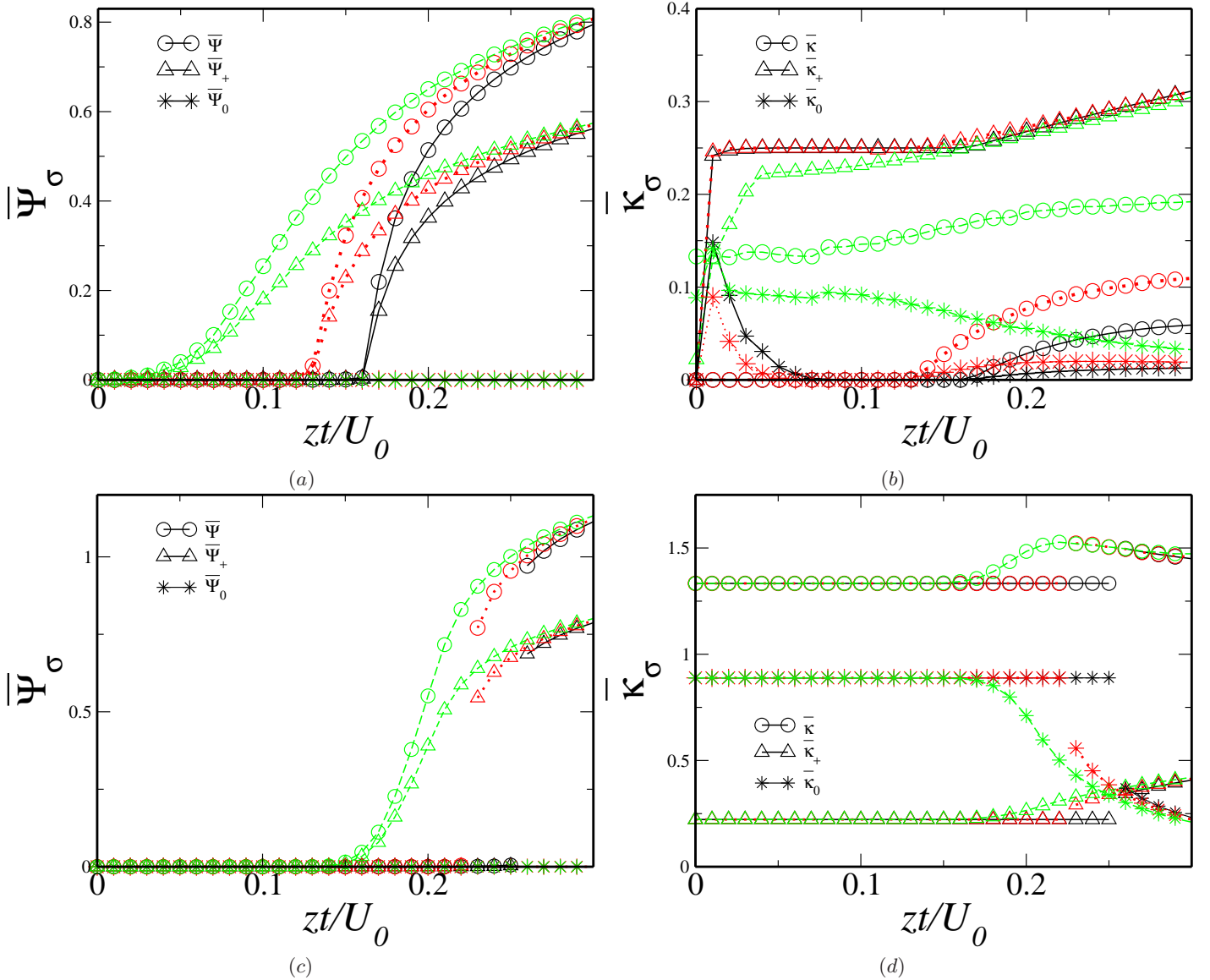

  \centerline{
    \hfill \psfig{file=psi_polar0.1_mu0.4.eps,width=0.5\textwidth}
    \hfill
    \hfill \psfig{file=k_polar0.1_mu0.4.eps,width=0.5\textwidth}
    \hfill}
    \centerline{\hfill $(a)$ \hfill\hfill  $(b)$ \hfill} 
     \centerline{
    \hfill \psfig{file=psi_polar0.1_mu1.4.eps,width=0.5\textwidth}
    \hfill
    \hfill \psfig{file=k_polar0.1_mu1.4.eps,width=0.5\textwidth}
    \hfill}
   \centerline{\hfill $(c)$ \hfill\hfill  $(d)$ \hfill} 
  \caption{(Color online) The variation of $\bar\psi_{\sigma}$ and
    $\bar\kappa_{\sigma}$ at $U_{2}/U_{0}=+0.1$ (antiferromagnetic)
    with disorder strengths $\Delta/U_{0}=0.3$ (dotted (red) lines)
    and 0.5 (dashed (green) lines) respectively and
    $\bar\psi_{+}=\bar\psi_{-}$, $\bar\kappa_{+}=\bar\kappa_{-}$ and
    $\bar\psi_{0}=\bar\kappa_{0}=0$. The pure case ($\Delta/U_{0}=0$)
    is included for comparison (solid (black) lines). At
    $\mu/U_{0}=0.4$, $\bar\psi$ and $\bar\kappa$ show continuous
    variations indicating a second order phase transition for MI lobes
    with $\rho=1$ [(a) and (b)]. While at $\mu/U_{0}=1.4$, $\bar\psi$
    and $\bar\kappa$ show continuous variation indicating a second
    order transition for MI lobes with $\rho=2$ at $\Delta/U_{0}$=0.5.
    [(c) and (d)].}
  \label{psik_polar}
\end{figure*}
\\ \\ \textbf{A. The behaviour of the SF order parameter and
  compressibility} \\ \indent In this section we present our results
of the averaged order parameter, $\bar{\psi}$ and compressibility,
$\bar{\kappa}$ to characterize the MI, BG and SF phases as a function
of tunneling strength $zt/U_{0}$. In the pure case, the $\bar{\psi}$
and $\bar{\kappa}$ undergo a transition from zero to finite values
indicating a direct MI-SF transition. That is, the system remains in
the MI phase as long as the tunneling strength, $t$ is well below the
critical tunneling strength $t_{c}$, where the latter is given by
\cite{Tsuchiya},
\begin{equation}
zt_{c}=(1/3)(U_{0}+2U_{2})[(2\rho+3)-\sqrt{4\rho^{2}+12\rho}]
\end{equation} 
While in the disordered case, the BG phase appears and it tries to
displace the MI phase making inroads for itself. The behaviours of
$\bar{\psi}$ and $\bar{\kappa}$ in both antiferromagnetic and
ferromagnetic regions is discussed below.\\\\ \indent (i)
Antiferromagnetic case: The spin dependent interaction, $U_{2}$ for
$^{23}Na$ is antiferromagnetic since the atomic scattering length
$a_{2}>a_{0}$. For $U_{2}/U_{0}>0$, the nature of MI-SF phase
transition is either of second or first order corresponding to the MI
lobes with odd or even local densities respectively, while for a
scalar Bose gas, it is always a second order phase transition.
\\ \indent In the Mott insulating phase, with an odd value for the
local density per site, the SF order parameter, $\bar\psi$ and the
compressibility, $\bar\kappa$ changes continuously from zero, thereby
indicating a second order MI-SF phase transition. However, the MI
phase with each site having even local density, the $\bar\psi$ and
$\bar\kappa$ show finite jumps indicating the first order MI-SF phase
transition \cite{Pai}, \cite{Kimura}.\\ \indent We have also found that
the three SF order parameters assume the values, such as
$\psi_{i+}$=$\psi_{i-}\ne 0$ and $\psi_{i0}$=0 in the MI phase making
the ground state only functions of $\psi_{i+}$ and $\psi_{i-}$ and
hence the equilibrium value of the total spin is $\langle {\textbf{S}}
\rangle^{2}$=0, which is responsible for making the SF phase a polar
state \cite{Pai}, \cite{Stefan}.  \\ \indent In order to visualize the
different kinds of phase transition corresponding to the MI lobes with
different occupation densities, we plot $\bar\psi$ and $\bar\kappa$ in
the pure and disordered cases. We choose the values of $\mu$ in such a
way so that we are at the MI lobes with local density $\rho_{i}$=odd
and even. At $\mu/U_{0}$=0.4, we are in the vicinity of the odd MI
lobe with the local density $\rho_{i}=1$ while for $\mu/U_{0}=1.4$,
the local density comes out as $\rho_{i}$=2 implying we are at the
even MI lobe.  \\ \indent In the pure case, we found that
corresponding to $U_{2}/U_{0}$=0.1, the system remains in the Mott
insulating phase with $\bar\psi$ and $\bar\kappa$ as zero till
$zt/U_{0}$=0.16 for the first odd MI lobe at $\mu/U_{0}$=0.4 and
$zt/U_{0}$=0.26 for the first even MI lobe at $\mu/U_{0}$=1.4, both of
which lie below the critical tunneling strength $zt_{c}/U_{0}$. Beyond
this critical value, the system goes to a superfluid phase with non
integer local densities, $\bar\rho$ and finite values of $\bar\psi$
[see Fig.\ref{psik_polar} (solid (black) lines)].  \\ \indent We shall now
include the on-site disorder, $\epsilon_{i}$ and see how it affects
the averaged SF order parameter and compressibility. The inclusion of
$\epsilon_{i}$ can equivalently be treated as a site dependent
chemical potential, $\mu_{i}$ ($\mu_{i}$=$\mu-\epsilon_{i}$). Since
the SF order parameter and local densities become explicitly site
dependent, it is impossible to determine the sharp value of the
tunneling strength, $zt$ for which a MI-BG transition can be
observed. The variation of averaged SF order parameter $\bar\psi$ and
$\bar\kappa$ for both the odd and even MI lobes for two different
disorder strengths are shown in Fig.[\ref{psik_polar}] at
$\Delta/U_{0}$=0.3 (dotted (red) lines) and $\Delta/U_{0}$=0.5 (dashed
(green) lines).
\begin{figure*}
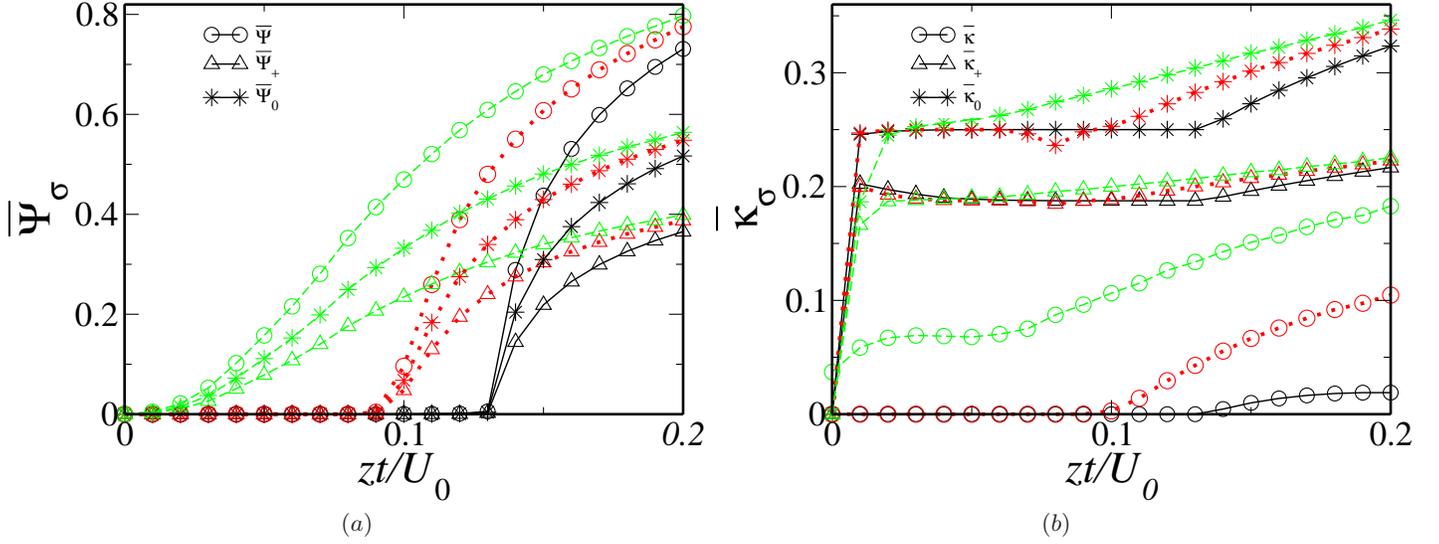

  \centerline{
    \hfill \psfig{file=psi_ferro0.2_mu0.4.eps,width=0.5\textwidth}
    \hfill
    \hfill \psfig{file=k_ferro0.2_mu0.4.eps,width=0.5\textwidth}
    \hfill}
   \centerline{\hfill $(a)$ \hfill\hfill  $(b)$ \hfill} 
  \caption{(Color online) The variation of $\bar\psi_{\sigma}$ and
    $\bar\kappa_{\sigma}$ at $U_{2}/U_{0}$=-0.2 (ferromagnetic) with
    disorder strengths $\Delta/U_{0}=0.3$ (dotted (red) lines) and 0.5
    (dashed (green) lines) respectively and
    $\bar\psi_{+}$=$\bar\psi_{-}$,
    $\bar\kappa_{+}$=$\bar\kappa_{-}$. The pure case
    ($\Delta/U_{0}=0$) is included for comparison (solid (black)
    lines). At $\mu/U_{0}=0.4$, $\bar\psi$ and $\bar\kappa$ show
    continuous variation indicating a second order phase transition
    for MI lobes with $\rho=1$ [(a) and (b)]}
  \label{psikferro}
\end{figure*}
\\ \indent For the odd MI lobes, we found that the MI region starts to
shrink due to appearance of the BG phase [see Figs.\ref{psik_polar}(a)
  and \ref{psik_polar}(b)] and the region spanned by the MI phase
gradually decreases at the larger value of the disorder strength. When
the disorder strength exceeds the critical value of disorder for the
respective MI lobes, the BG phase completely destroys the MI phase,
resulting in only the existence of the BG and SF phases as pointed out
earlier. Thus at $\Delta/U_{0}$=0.5, the BG region extends till
$zt/U_{0}\simeq 0.072$ as perceived from the vanishing of the average
SF order, $\bar\psi$ below 0.1 and finite average compressibility,
$\bar\kappa$.  \\ \indent We also study the behavior of even MI lobes
in the antiferromagnetic case at different disorder strengths. In the
pure case, we choose $\mu/U_{0}$=1.4 and we found that both $\bar\psi$
and $\bar\kappa$ show jumps from zero and a constant value
respectively to distinct finite values signaling the first order
phase transition for the even MI lobes [see Figs.\ref{psik_polar}(c)
  and \ref{psik_polar}(d) (solid (black) lines)]. In presence of
disorder, the MI state starts to diminish due to the appearance of the
BG phase and the MI-SF phase transition continues to be first order
for $\Delta/U_{0}$=0.3 [see Figs.\ref{psik_polar}(c) and
  \ref{psik_polar}(d) (dotted (red) lines)]. While at
$\Delta/U_{0}$=0.5 [see Figs.\ref{psik_polar}(c) and
  \ref{psik_polar}(d) (dashed (green)lines)], the MI state still
survives but the MI-SF transition becomes a second order phase
transition. \\ \indent We have also investigated the behaviour of the
individual SF order parameter and local density components at
different disorder strengths at both values of chemical potential,
namely $\mu/U_{0}$=0.4 and 1.4. At $\Delta/U_{0}$=0, it is found that
the SF order parameters behave as, $\psi_{+}$=$\psi_{-}$=$\psi_{0}$=0
with the occupation densities becoming $\rho_{+}$=$\rho_{-}\simeq 0.5$
and $\rho_{0}\simeq 0$ at $\mu/U_{0}$=0.4 and
$\rho_{+}$=$\rho_{-}$=$\rho_{0}$=2/3 at $\mu/U_{0}$=1.4 in the MI
phase. While $\psi_{+}$=$\psi_{-}\ne 0$, $\psi_{0}$=0 and $\rho_{+}$,
$\rho_{-}$, $\rho_{0}$ behave similar to that of $\rho$ in the SF
phase satisfying the condition
$\rho$=$\sum\nolimits_{\sigma}\rho_{\sigma}$ and correspondingly
$\bar\psi_{\sigma}$ and $\bar\kappa_{\sigma}$ at $\Delta/U_{0}$=0.3,
0.5 are shown in Fig.[\ref{psik_polar}]. The compressibility,
$\bar\kappa_{\sigma}$ is constant in the MI phase and gradually
increases in the SF phase for even MI lobes. \\ \indent The above
scenario is equivalent to the MI phase with fractional occupation
densities for an individual spinor components, although the summed
over spinor components yields an integer. \\\\ \indent
(ii)Ferromagnetic case: For $^{87}Rb$ atoms, the experimental data
yield $a_{2}<a_{0}$ for the scattering lengths in the respective
channels for which the spin dependent interaction $U_{2}\le 0$ and the
spin-1 ultracold atoms show similar behavior as that of scalar or
spinless Bose gas where the MI-SF phase transition is now a second
order one. \\ \indent In the ferromagnetic regime, all the three
components of the SF order parameter are non zero in the SF phase and
the ground state is now functions of $\psi_{i+}$, $\psi_{i-}$,
$\psi_{i0}$. In the pure case, we found that $\psi_{i+}=\psi_{i-}$ and
$\psi_{i0}=\sqrt{2}\psi_{i+}$ and the total spin operator $\langle
{\textbf{S}} \rangle^{2}$ in the superfluid phase now only takes the
value of unity which confirms that we have a ferromagnetic superfluid
state \cite{Pai}, \cite{Stefan}.  \\ \indent The variation of
$\bar\psi$ and $\bar\kappa$ are shown in Figs.[\ref{psikferro}(a) and
  \ref{psikferro}(b)] for both pure and disordered cases at
$U_{2}/U_{0}=-0.2$ and $\mu/U_{0}=0.4$. When $\Delta/U_{0}=0$, the
local density, $\rho=1$ and $\bar\psi$, $\bar\kappa$ are zero till
$zt/U_{0}$=0.13 ($<zt_{c}$) indicating the system is in the MI
regime. As $zt/U_{0}$ increases, the system smoothly goes toward the
SF regime with finite $\bar\psi$ and $\bar\kappa$ and non integer
local densities [see Figs.\ref{psikferro}(a) and \ref{psikferro}(b)
  (solid (black) lines)].  \\ \indent As disorder is introduced, the
BG phase intervenes in between the MI and SF phases. Thus the width of
the MI phase at a disorder strength, $\Delta/U_{0}$=0.5 [see
  Figs.\ref{psikferro}(a) and \ref{psikferro}(b) (dashed (green)
  lines)] becomes much lesser as compared to that of
$\Delta/U_{0}$=0.3 [see Figs.\ref{psikferro}(a) and
  \ref{psikferro}(b)(dotted (red) lines)]. From Fig.[\ref{psikferro}],
we see that at $\Delta/U_{0}$=0.5, the region spanned by the MI phase
is completely occupied by the BG phase, resulting in only the survival
of the BG and the SF phases owing to similar reasons as that of the
antiferromagnetic case discussed earlier. \\ \indent We also study the
behavior of the individual SF order parameter and local density
components at different disorder strengths at $\mu/U_{0}$=0.4.  At
$\Delta/U_{0}$=0, $\psi_{+}$=$\psi_{-}$=$\psi_{0}$=0 and
$\rho_{+}$=$\rho_{-}\simeq 0.25$ and $\rho_{0}\simeq 0.5$. The
compressibility, $\bar\kappa_{\sigma}$ is constant in the MI phase and
corresponding $\bar\psi_{\sigma}$ and $\bar\kappa_{\sigma}$ at
$\Delta/U_{0}$=0.3, 0.5 are shown in Figs.[\ref{psikferro}(a)and
  \ref{psikferro}(b)].  \\ \\ \textbf{B. Indicators of MI, BG and SF
  phases} \\ \indent In the previous section we have studied the
effects of onsite disorder on the averaged SF order parameter and the
compressibility. Since we have averaged the SF order parameter and the
compressibility, it becomes very difficult to locate the precise
parameter values for a MI-BG and BG-SF transition, owing to the fact
that the $\psi_{i}$ and the $\rho_{i}$ are explicitly site
dependent. Thus at the transition point there are some sites with zero
SF order parameter and integer occupation densities, while other sites
may have zero SF order parameter and non integer occupation densities.
\\ \indent In order to circumvent this difficulty, we define a
measurable quantity, $\chi$ which is defined as,
\begin{equation}
\chi=\frac{Sites\hspace{1mm} with\hspace{1mm} \psi_{i}\ne
  0\hspace{2mm}and \hspace{1mm}\rho_{i}\ne
  integer}{Total \hspace{1mm}numbers\hspace{1mm} of\hspace{1mm} sites}
\label{chi}
\end{equation}
\indent In a similar manner, we can also define the $\chi_{\sigma}$
involving $\psi_{i\sigma}$ and $\rho_{i\sigma}$ which will aid in
the study of the individual spinor components.
\begin{figure*}
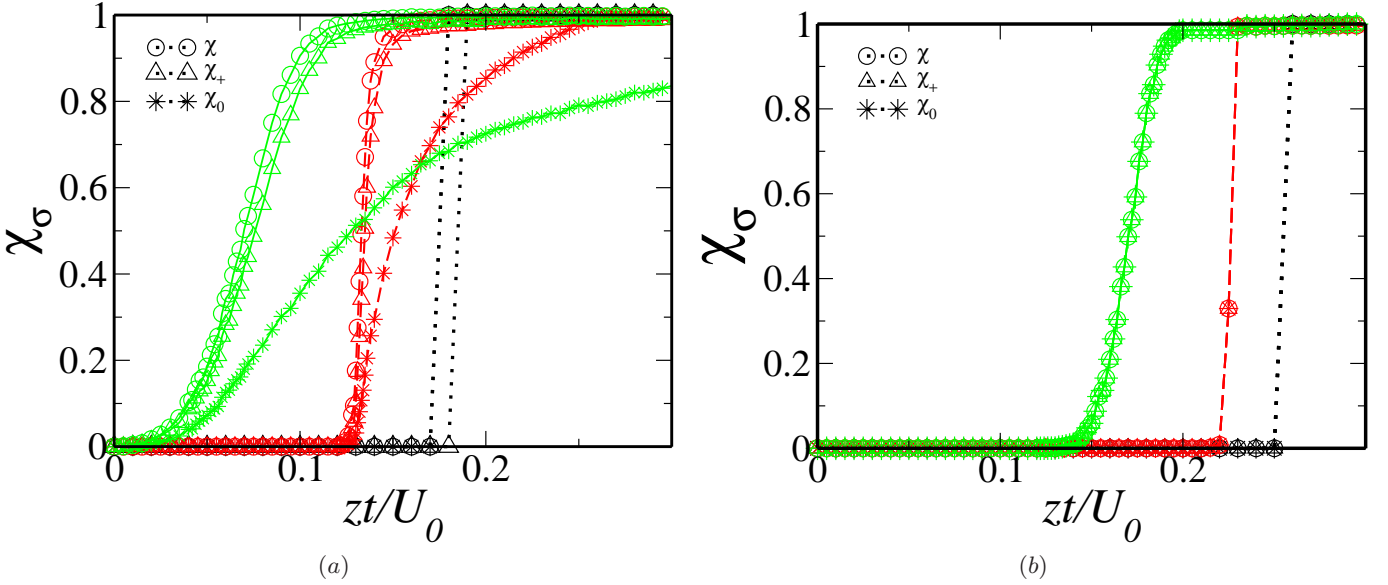

  \centerline{ \hfill
    \psfig{file=chi_polar_0.1_mu0.4.eps,width=0.48\textwidth} \hfill
    \hfill \psfig{file=chi_polar_0.1_mu1.4.eps,width=0.48\textwidth}
    \hfill} \centerline{\hfill $(a)$ \hfill\hfill $(b)$ \hfill}
  \caption{The variation of $\chi_{\sigma}$ in the antiferromagnetic
    case with $U_{2}/U_{0}$=+0.1 for both the odd and even MI lobes
    corresponding to the pure case $\Delta/U_{0}$=0.0 (dotted (black)
    lines) and for disordered values, $\Delta/U_{0}$=0.3 (dashed (red)
    lines) and 0.5 (solid (green) lines) respectively. $\chi_{\sigma}$
    for odd MI lobe at $\mu/U_{0}$=0.4 is shown in (a) while for the
    even MI lobes at $\mu/U_{0}$=1.4 in (b).}
\label{chipolar}
\end{figure*}
\\ \indent The main reason to define $\chi$ is to distinguish between
the MI, BG and SF phases where depending upon the value of $\chi$, we
will be able to easily characterize the three different types of
phases.\\ \indent At this point, one can ask, what will be the value
of $\psi_{i}$ and $\rho_{i}$ which can define the MI or the SF
phase. Since we see from the behaviour of $\bar\psi$ and $\bar\kappa$
in the pure case that the system will remain in the MI phase till
$zt<zt_{c}$ having vanishingly small $\psi_{i\sigma}$ (values below
$O(10^{-2})$) and the occupation density $\rho_{i}=N\pm \delta$ where
$N=0,1,2,3...$ and $\delta$ is of $O(10^{-3})$. These values are set
for numerical convergence of the parameters corresponding to the MI
phase. Again as earlier, we include discussion of the
antiferromagnetic and ferromagnetic regimes one by one.  \\\\ \indent
(i) Antiferromagneic case: For antiferromagnetic interaction with
$U_{2}/U_{0}$=0.1, we study the the variation of $\chi$ with tunneling
strength, $zt/U_{0}$ in both pure and disordered cases at two
different values of $\mu/U_{0}$ corresponding to the odd and even MI
lobes respectively as shown in Fig.[\ref{chipolar}].  \\ \indent In
the disorder free case for $\mu/U_{0}$=0.4, we see that $\chi$ makes
direct transition from 0 to 1 emphasizing an unhindered MI-SF
transition. When $zt<zt_{c}$, the system is in MI phase, that is all
sites have $\psi_{i}$=0 and $\rho_{i}$=integer and thus $\chi$ takes
the value 0. While for $zt>zt_{c}$, the system becomes a SF phase with
$\psi_{i}\ne 0$ and $\rho_{i}\ne$ integer and thereby $\chi$ assumes a
value 1 [see Fig.\ref{chipolar}(a) (dotted (black) lines)]. As soon as
disorder is included, the direct transition of $\chi$ from 0 to 1 is
prohibited indicating a gradual increase due to the presence of the BG
phase.  \\ \indent In presence of disorder, that is at
$\Delta/U_{0}$=0.3 [see Fig.\ref{chipolar}(a) (dashed (red) lines)],
$\chi$ remains zero till $zt/U_{0}=0.133$, beyond which $\chi$ starts
to increase indicating that some sites with $\rho_{i}$=integer in the
MI regime start evolving and the system move towards the BG regime
with $\rho_{i}\ne$ integer. Further, the gradual increase of the slope
of $\chi$ refers the intrusion of the BG phase into a territory which
used to be the MI phase without disorder. As disorder strength
increases, that is at $\Delta/U_{0}$=0.5 [see Fig.\ref{chipolar}(a)
  (solid (green) lines)], we see that $\chi$ only takes finite values
and ultimately go unity as a function of $zt/U_{0}$, indicating that
the region spanned by the BG phase continues to increase and results
in vanishing of the MI phase. Thus at $\Delta/U_{0}$=0.5, the system
only consists of the BG and SF phases as expected since the disorder
strength $\Delta/U_{0}$ exceeds the energy gap for the first odd MI
lobe ($\rho$=1).  \\ \indent It is now quite interesting to study the
behavior of $\chi_{\sigma}$ incorporating the effects of disorder. In
order to study $\chi_{\sigma}$, we use $\psi_{i\sigma}$ and
$\rho_{i\sigma}$ and remember that they satisfy the same condition as
the total SF order parameter, $\psi_{i}$ and the local density,
$\rho_{i}$ for rendering the MI and the SF phases respectively.
\\ \indent At $\mu/U_{0}$=0.4, the system is in the MI regime since
the SF order parameter $\psi_{i}$ (=$\sqrt{\psi^{2}_{i\sigma}}$) is
small and the local densities are as, $\rho_{i}$=1,
$\rho_{i+}$=$\rho_{i-}\simeq 0.5$, $\rho_{i0}\simeq O(10^{-3})$. Hence
following the definition of $\chi$, we compute $\chi_{\pm}$ by
suitably modifying the numerator of the Eq.(\ref{chi}) with
$\psi_{i\pm}\ne 0$ and $\rho_{i\pm}\ne0.5$ respectively. In
Fig.[\ref{chipolar}(a)], we have plotted $\chi_{\pm}$ with
$zt/U_{0}$. We find that $\chi_{+}$=$\chi_{-}$ where they show same
behaviour as that of $\chi$ in the clean and disordered
cases.\\ \indent In the antiferromagnetic case, $\psi_{i0}$ is always
zero and $\rho_{i0}$ behaves as $\rho_{\pm}$ corresponding to the MI
and SF regimes respectively. So for determining $\chi_{0}$, we set the
condition $\psi_{i0}$=0 and $\rho_{i0}> 0.001$ in the numerator of
Eq.(\ref{chi}). Its variation with respect to $zt/U_{0}$ is also shown
in the Fig.[\ref{chipolar}(a)].  \\ \indent Similar to the discussion
above, at $\mu/U_{0}$=1.4, in the MI regime, the local densities yield
an even MI lobe with $\rho_{i}$=2, $\rho_{i+}$=
$\rho_{i-}$=$\rho_{i0}$=$2/3\pm \delta$. So we compute $\chi_{\pm}$ by
replacing the numerator of the Eq.(\ref{chi}) with $\psi_{i\pm}\ne 0$
and $\rho_{i\pm} \ne 0.667\pm\delta$. For computing $\chi_{0}$, we set
the condition $\psi_{i0}$=0 and $\rho_{i0}\ne 0.667\pm \delta$
respectively. The variation of $\chi$ and $\chi_{\sigma}$ with
different disorder strengths are shown in
Fig.[\ref{chipolar}(b)].\\ \indent From the behaviour of $\chi$, we
see that at $\Delta/U_{0}$=0.3 [see Fig.\ref{chipolar}(b)(dashed (red)
  lines)], although there is little value in between 0 and 1, but the
region occupied by the MI phase shrinks as compared with the pure case
[see Fig.\ref{chipolar}(b) (dotted (black) lines)], owing to the
emergence of the BG phase.  For $\Delta/U_{0}$=0.5 [see
  Fig.\ref{chipolar}(b) (solid (green) lines)], the BG phase appears
as $\chi$ takes finite values in between 0 and 1, while the MI phase
still exists since the critical disorder strength, $\Delta_{c}/U_{0}$
for the $\rho$=2 MI lobe is a much higher as discussed in the previous
section. Components of $\chi_{\sigma}$ show similar kind of behavior
as that of $\chi$ in presence of disorder.  \\ \\\indent
(ii)Ferromagnetic case: For ferromagnetic interactions with
$U_{2}/U_{0}$=-0.2 at $\mu/U_{0}$=0.4, the variation of
$\chi_{\sigma}$ with tunneling strength $zt/U_{0}$ is shown in
Fig.[\ref{chiferro}]. In the ferromagnetic case, since the spin-1
ultracold atoms shows similar phase diagram as that of a scalar Bose
gas (spinless), so there is no distinction between the odd and the
even MI lobes, and thus showing results involving either will suffice.
\\ \indent At $\mu/U_{0}$=0.4, the MI lobes has local density
$\rho_{i}$=1 and $\chi$ is zero till $zt/U_{0}$=0.133. Beyond this
value, the system goes to the SF phase with $\chi$=1 in the pure case
[see Fig.\ref{chiferro} (dotted (black) lines)]. In presence of
disorder, that is at $\Delta/U_{0}$=0.3 [see Fig.\ref{chiferro}
  (dashed (red) lines)], $\chi$ remains zero till $zt/U_{0}$=0.085,
beyond which $\chi$ starts to increase and at $\Delta/U_{0}$=0.5 [see
  Fig.\ref{chiferro} (solid (green) lines)] $\chi$ only takes the value
between 0 and 1 indicating that the system only consists of the BG and
SF phases. In the latter case, disorder completely annihilates the MI
phase because the disorder strength crosses the critical disorder
value, $\Delta_{c}/U_{0}$=0.4.  \\ \indent At $\mu/U_{0}$=0.4, in the
MI regime, the SF order parameter, $\psi_{i}$
(=$\sqrt{\psi^{2}_{i\sigma}}$) is small and the local densities are
given by $\rho_{i}$=1, $\rho_{i+}$=$\rho_{i-}$=$0.25\pm \delta$,
$\rho_{i0}$=$0.5\pm \delta$. We compute the $\chi_{\pm,0}$ by
modifying the Eq.(\ref{chi}) with $\psi_{i\pm,0}\ne 0$ and
$\rho_{i\pm,0}$ respectively, since in the SF regime, all the three SF
spinor components are finite. In Fig.[\ref{chiferro}] we have plotted
$\chi_{\pm,0}$ as a function of $zt/U_{0}$ which show same behaviour
as that of $\chi$ in both the clean and disordered cases.
\begin{figure}
  \centerline{
    \hfill \psfig{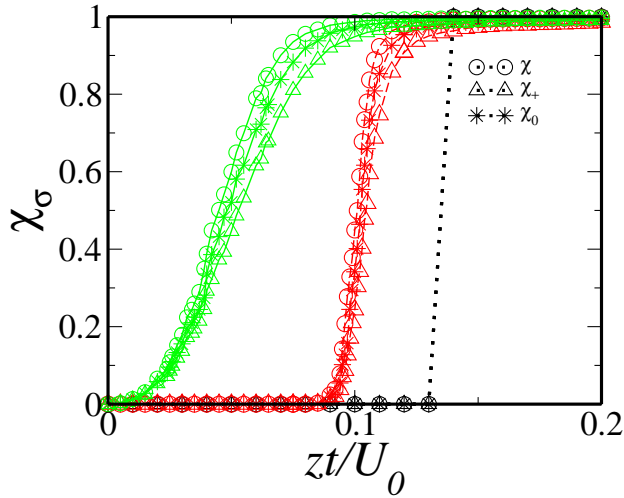}
    \hfill}
  \caption{The variation of $\chi_{\sigma}$ in the ferromagnetic case
    with $U_{2}/U_{0}$=-0.2 corresponding to the pure case
    $\Delta/U_{0}$=0.0 (dotted (black) lines) and for disordered
    values, $\Delta/U_{0}$=0.3 (dashed (red) lines) and 0.5 (solid
    (green) lines) respectively.}
\label{chiferro}
\end{figure}
\\ \\ \textbf{C. Percolation analysis and finite size scaling}
\begin{figure*}
  \centerline{
    \hfill \psfig{file=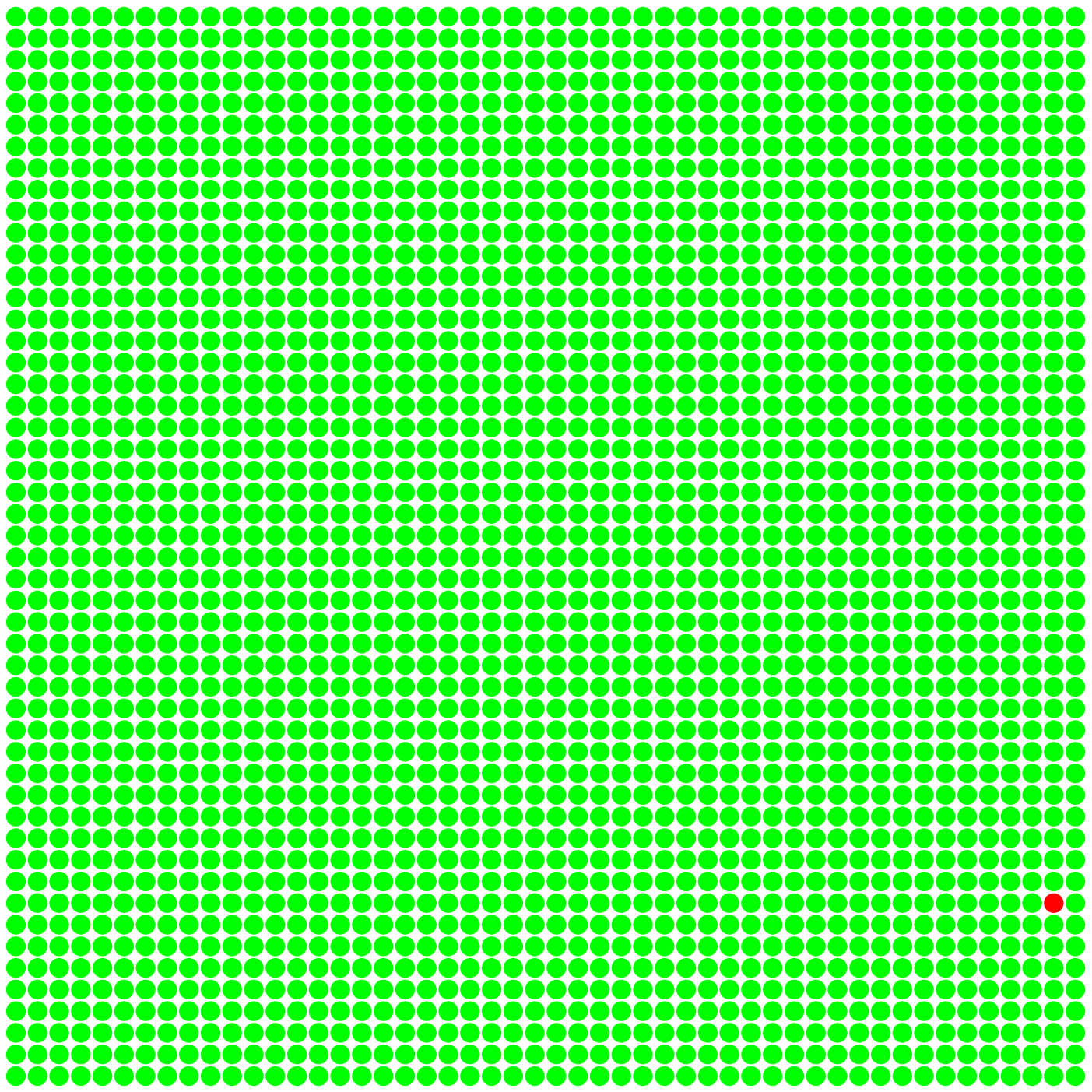,width=0.31\textwidth}
    \hfill
    \hfill \psfig{file=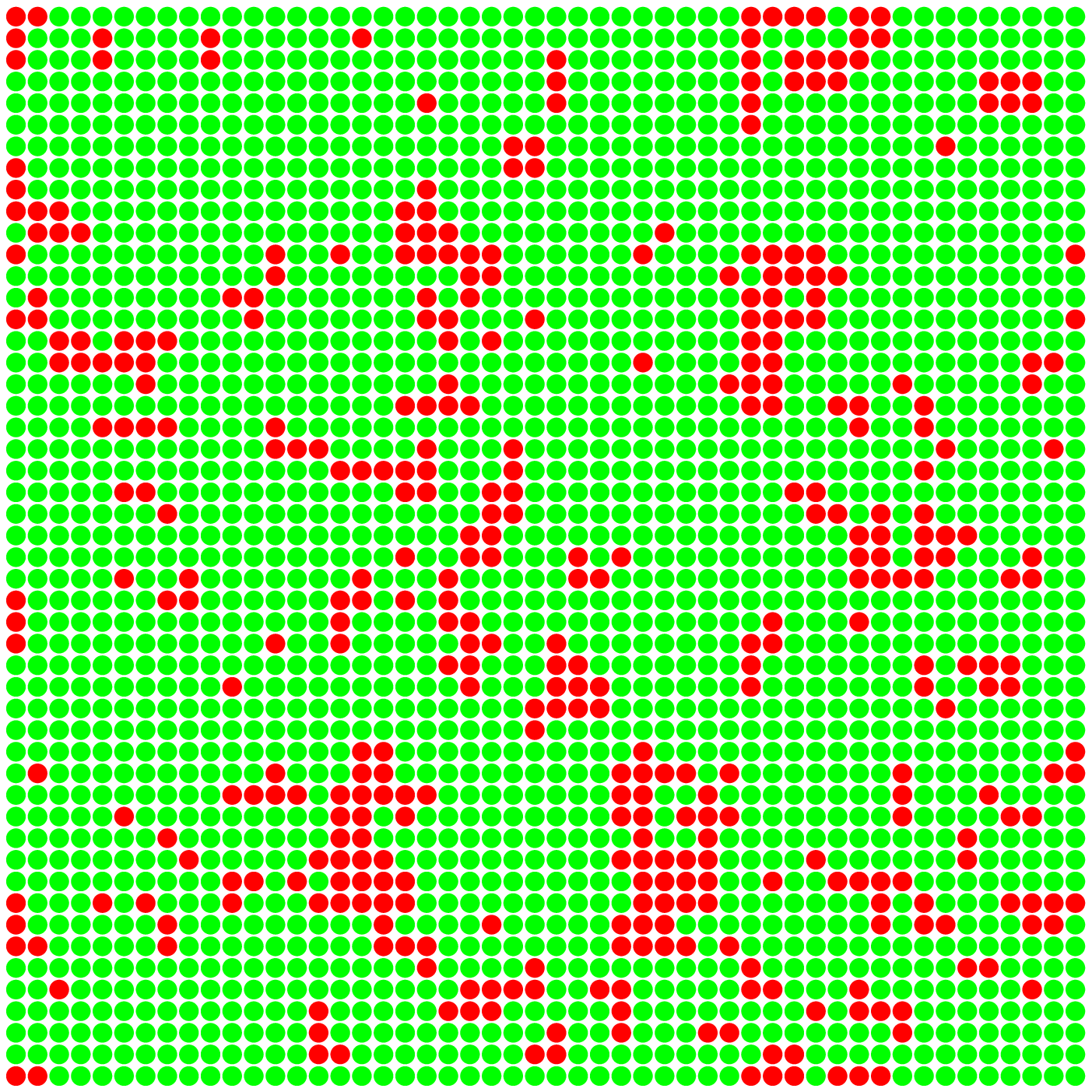,width=0.31\textwidth}
    \hfill
    \hfill \psfig{file=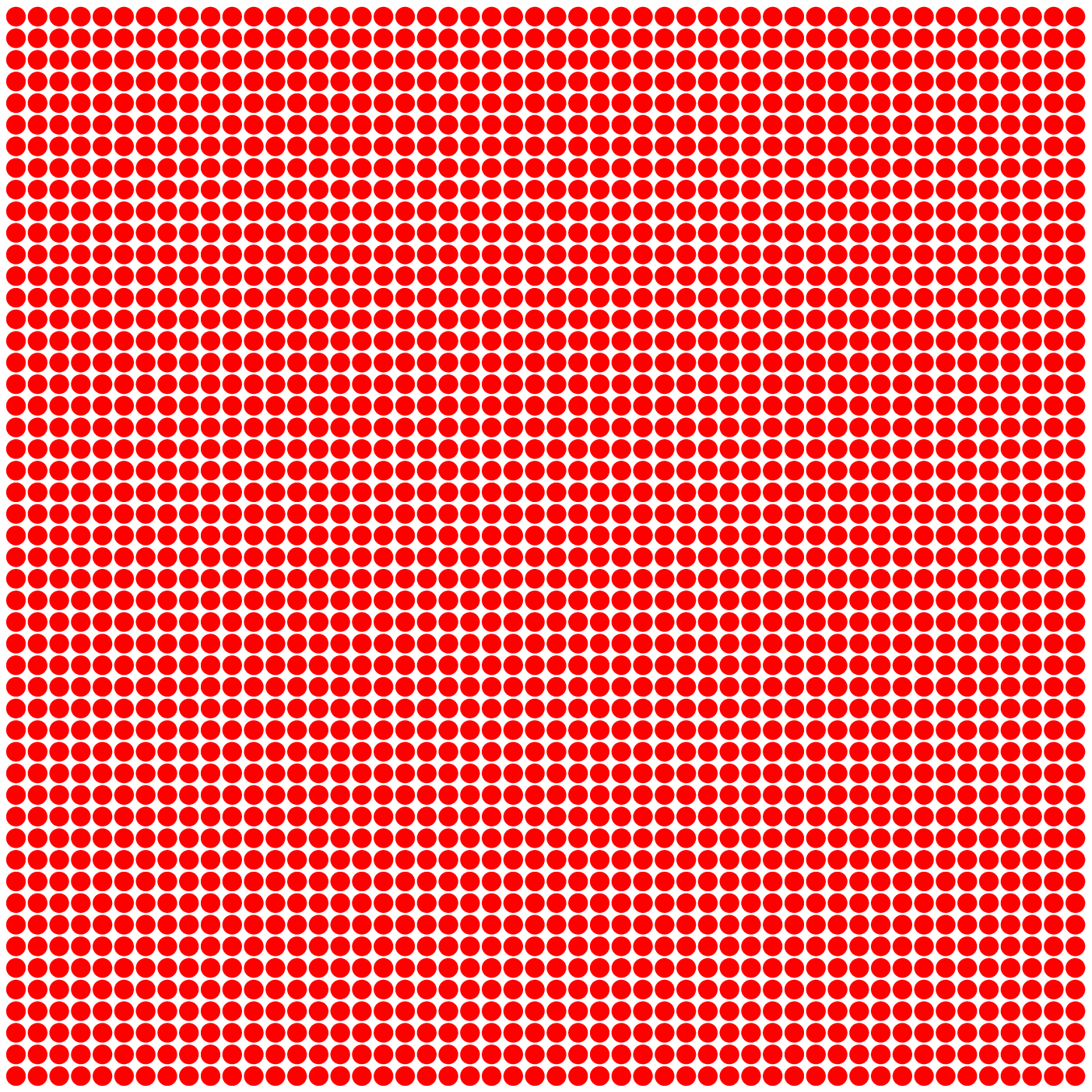,width=0.31\textwidth}
    \hfill}
    \vspace{3mm}
     \centerline{
    \hfill \psfig{file=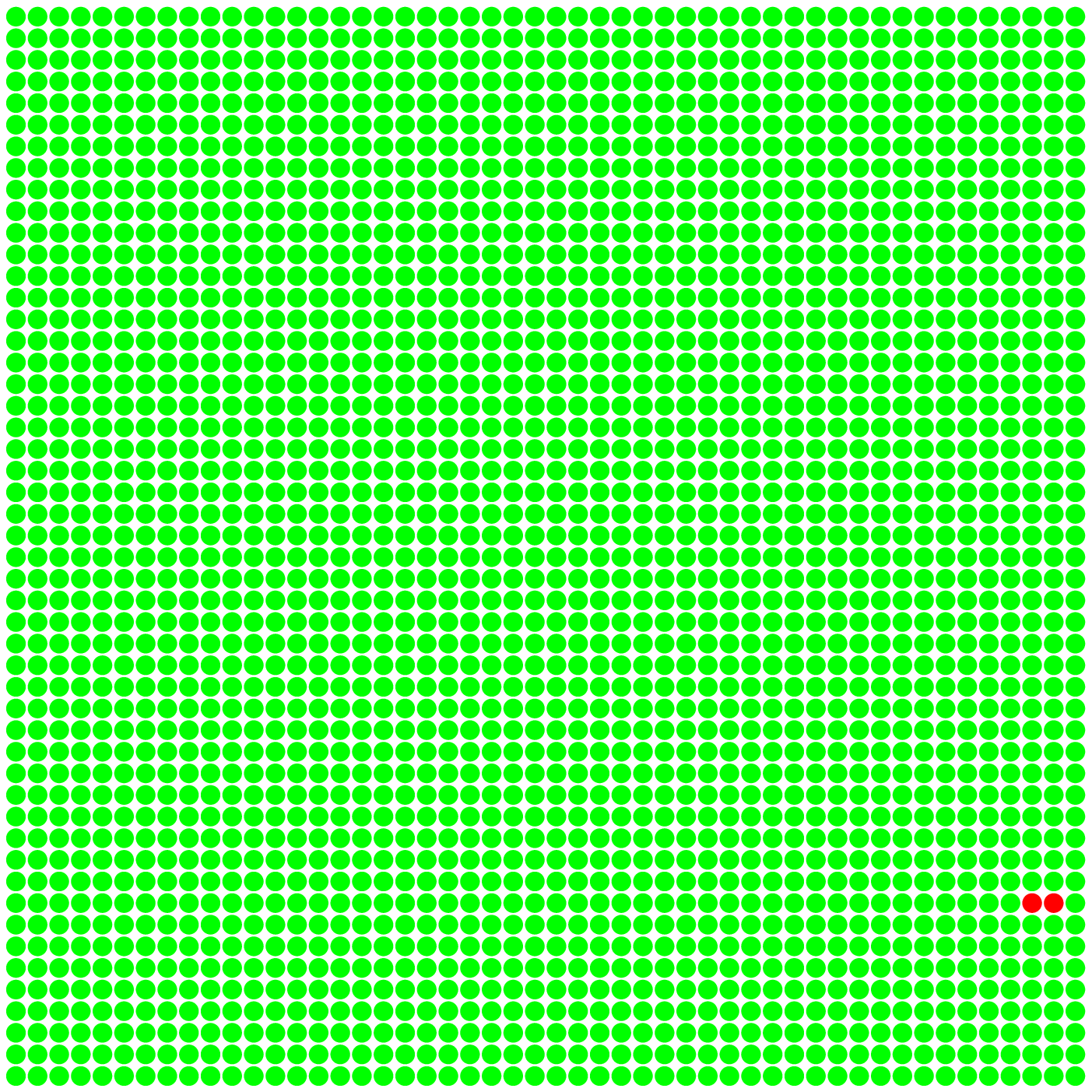,width=0.31\textwidth}
    \hfill
    \hfill \psfig{file=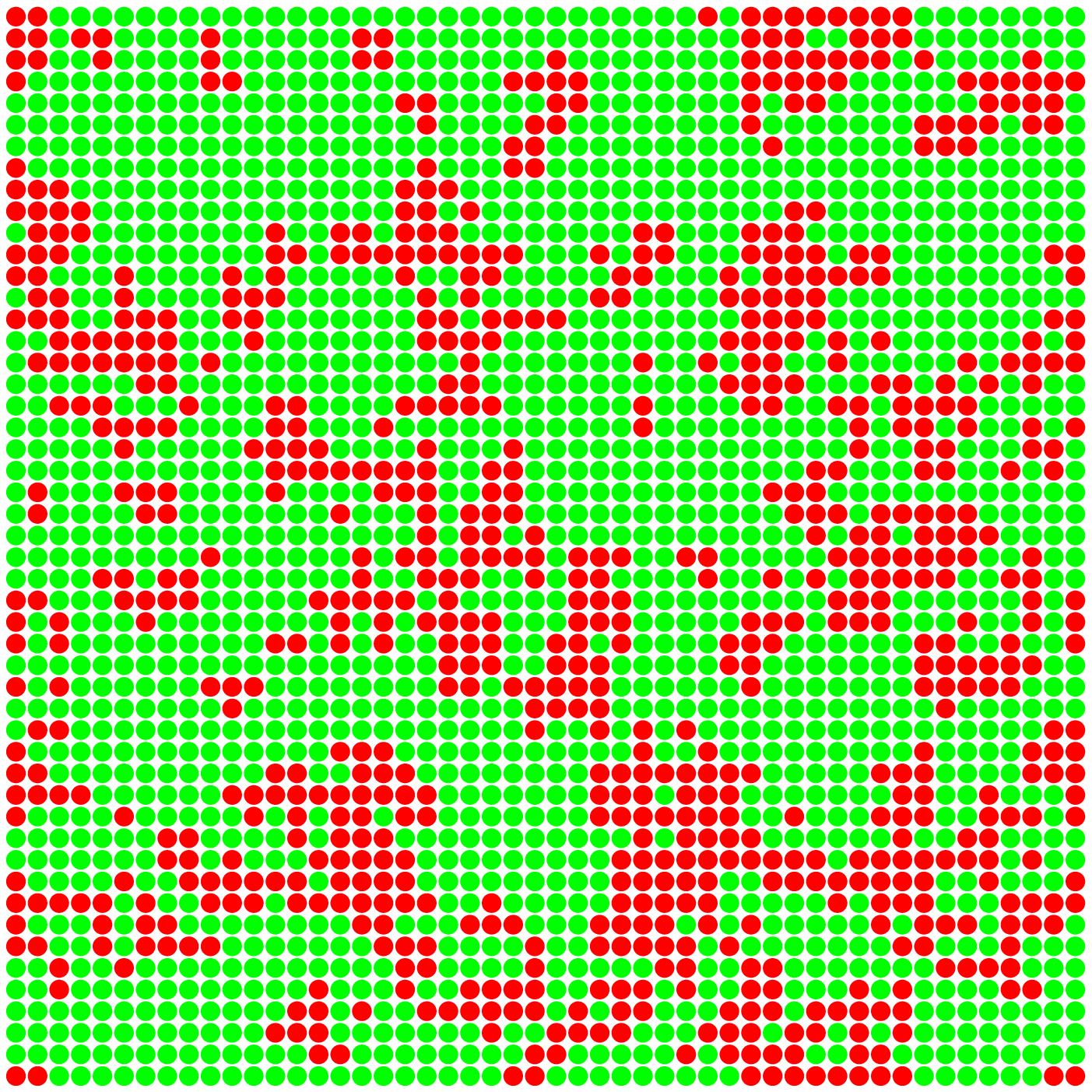,width=0.31\textwidth}
    \hfill
    \hfill \psfig{file=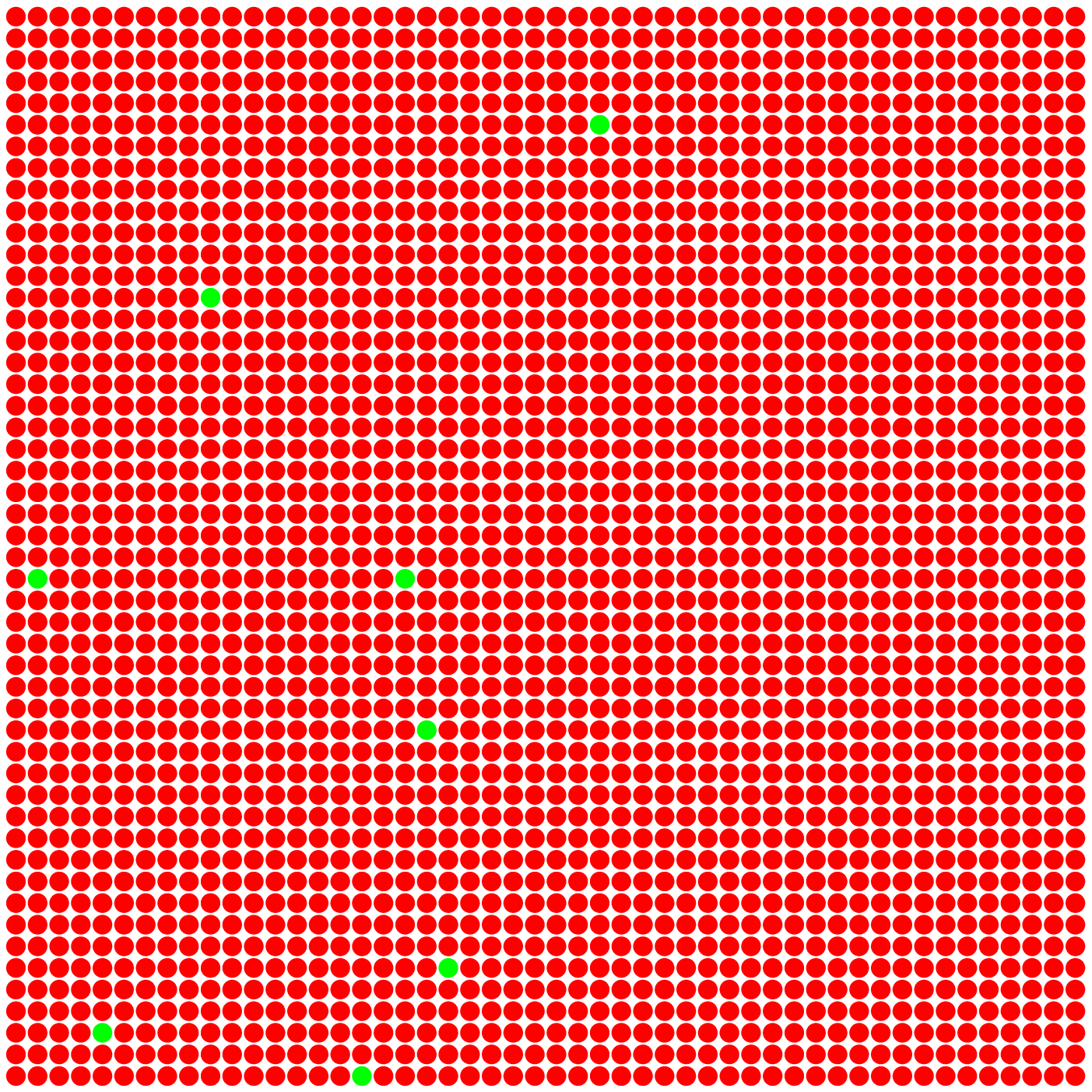,width=0.31\textwidth}
    \hfill}
    \centerline{$(a) MI$\hspace{5cm}  $(b) BG$ \hspace{5cm} $(c) SF$} 
  \caption{(color online) The real space plots of SF order parameter,
    $\psi_{i}$ (first row) and the occupation densities, $\rho_{i}$
    (second row) for lattice size $L\times L$=$50\times 50$ at
    $\Delta/U_{0}$=0.5 with $U_{2}/U_{0}$=+0.1 and $\mu/U_{0}$=0.4
    corresponding to three different phases that is the MI, BG and SF
    phases of the system respectively. The light green circles (light
    shades) represent the zero SF order parameter and integer
    densities and deep red circles (deep shades) represent the non
    zero SF order parameter and non integer densities. The parameter
    value is at $zt/U_{0}$=0.001 in the MI phase (a) and
    $zt/U_{0}$=0.06 in the BG phase (b) and $zt/U_{0}$=0.3 for the SF
    phase (c) respectively.}
\label{realspace}
\end{figure*}
\\ \indent At this point we will try to figure out one of our earlier
questions that was posed in this regard, whether is it possible to
locate the transition point for the MI-BG transition in the presence
of disorder with confidence. Further what will be the extent of the BG
phase, that is, whether the whole region is a BG phase or there is
certain amount of SF phase still left in presence of disorder. To
answer the questions we have to really look at the concept of the
appearance of percolating cluster of sites with non integer occupation
densities and finite SF order parameter. In order to understand what
do we mean by a percolating cluster, we have shown the typical real
space plots of $\psi_{i}$ and $\rho_{i}$ for three different values of
$zt/U_{0}$ corresponding to the three different phases of the system
in Fig.[\ref{realspace}]. The pair of figures shown in each column
depicts MI [see Fig.\ref{realspace}(a)], BG [see Fig.\ref{realspace}(b)] and
SF [see Fig.\ref{realspace}(c)] phases. The top and bottom plots in each
of them illustrate the values of SF order parameter, $\psi_{i}$ and
the local density $\rho_{i}$ respectively in a square lattice of size
$L\times L$=$50\times50$ for a single realization of the disorder. The
parameter corresponding to the plots are $\Delta/U_{0}$=0.5 and
$\mu/U_{0}$=0.4 in the antiferromagnetic case
($U_{2}/U_{0}$=+0.1). The light green circles are the indicators for
the zero SF order parameter and the integer densities, while the deep
red circles denote non zero SF order parameter and non integer
densities. The $\psi_{i}$ and $\rho_{i}$ for the MI phase at
$zt/U_{0}$=0.001 and for the BG phase at $zt/U_{0}$=0.06 and for the
SF phase at $zt/U_{0}$=0.3.
\begin{figure*}
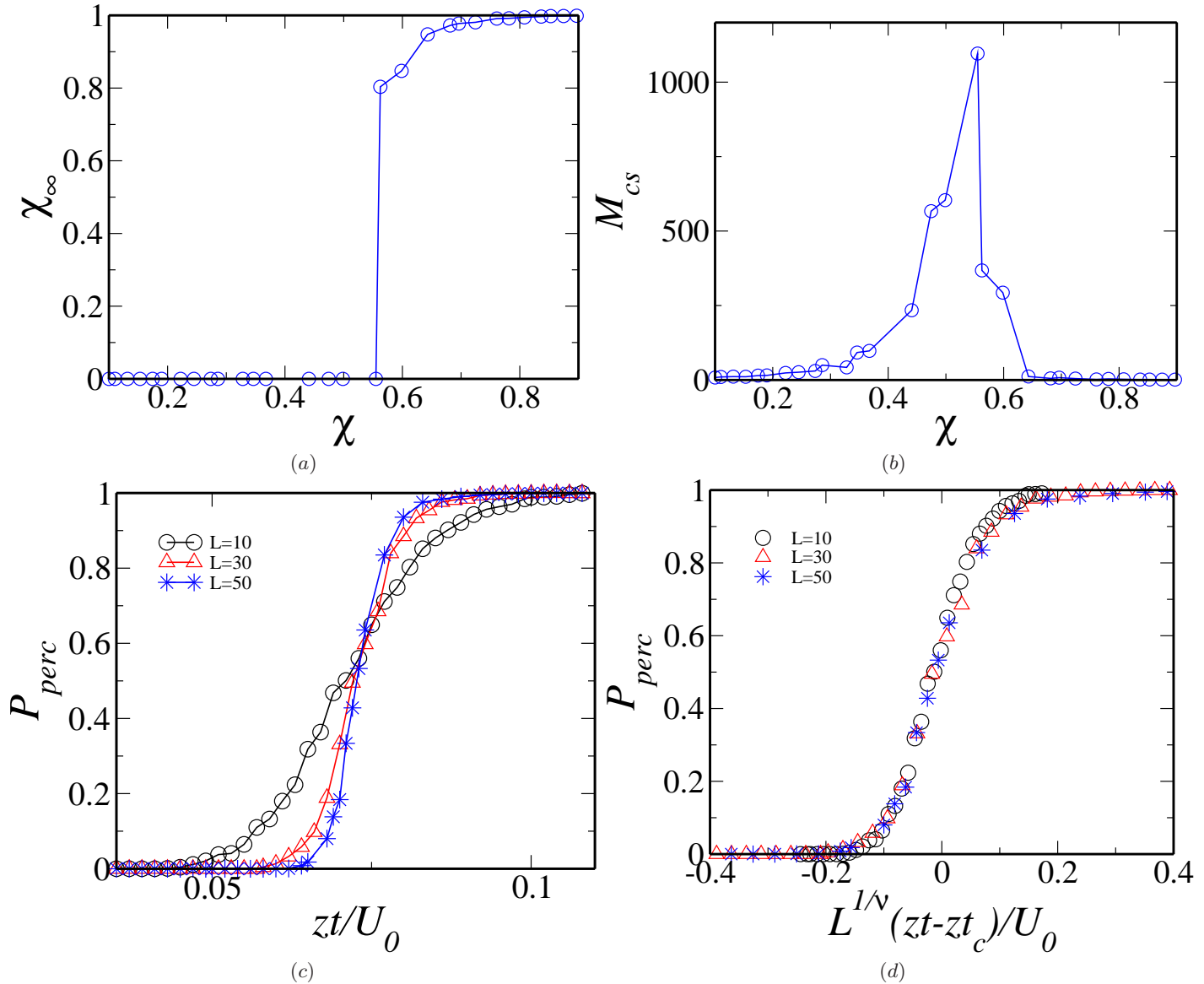

  \centerline{
    \hfill \psfig{file=chi_inftyL128.eps,width=0.48\textwidth}
    \hfill
    \hfill \psfig{file=M_cs_L128.eps,width=0.5\textwidth}
    \hfill}
    \centerline{\hfill $(a)$ \hfill\hfill  $(b)$ \hfill} 
    \centerline{
    \hfill \psfig{file=P_perc_polar.eps,width=0.5\textwidth}
    \hfill
    \hfill \psfig{file=finitesize_polar.eps,width=0.5\textwidth}
    \hfill}
   \centerline{\hfill $(c)$ \hfill\hfill  $(d)$ \hfill} 
  \caption{(Color online) The percolation probability, $\chi_{\infty}$
    and the mean cluster size, $M_{cs}$ corresponding to a lattice
    size $L\times L$=$128\times 128$ at $\Delta/U_{0}$=0.5 for the
    antiferromagnetic case with $U_{2}/U_{0}$=+0.1 and $\mu/U_{0}$=0.4
    are shown in (a) and (b). The percolation probability,
    $P_{perc}$ corresponding to different lattice size $L$ with system
    parameter, $\eta$=$zt/U_{0}$ and finite size scaling are shown in
    (c) and (d). The critical tunneling strength $\eta_{c}$ where
    $P_{perc}$ for different $L$ values intersect at
    $\eta_{c}$=$zt_{c}/U_{0}$=0.073.}
\label{finitesize}
\end{figure*}
\\ \indent In the clean state, the system splits into two, the MI
islands with all sites having zero $\psi_{i}$ and integer densities
$\rho_{i}$ as shown by light dots in Fig.[\ref{realspace}(a)] till
$zt/U_{0}<zt_{c}/U_{0}$ and the SF islands with finite $\psi_{i}$ and
non integer densities $\rho_{i}$ at $zt/U_{0}>zt_{c}/U_{0}$ separated
by a sharp boundary. As soon as disorder is introduced, the BG phase
tries to mix two islands by removing the boundary between them. So the
sites with zero $\psi_{i}$ and integer densities attempt to evolve
towards zero $\psi_{i}$ and non integer densities as shown in
Fig.[\ref{realspace}(b)] (BG phase), which finally percolates for the
first time towards a state with non zero $\psi_{i}$ and non integer
densities Fig.[\ref{realspace}(c)], the latter being the SF phase.
\\ \indent To extract different phases of the system, we shall
concentrate on the appearance of the SF clusters which percolate
throughout the entire lattice for the first time. In the MI states,
all sites have integer particle densities and hence there will be no
SF cluster, while in the BG phase, some of the sites have non integer
densities and thus the SF clusters will be trapped by other clusters
with sites having integer densities.  Finally with the increase in the
tunneling strength, the SF clusters will start to percolate throughout
the lattice resulting in a SF phase for the system. Thus the BG region
can be identified in between the MI and the SF phases where SF
clusters exist, however do not percolate across the lattice.
\\ \indent When a SF cluster percolates through an infinite lattice it
is known as an infinite cluster, while when it percolate through a
finite lattice it is known as a percolating or spanning cluster. Since
we are dealing with a finite size of the system for our numerical
work, we shall aim to find a spanning cluster using HK algorithm
\cite{Hoshen}.\\ \indent Whether a SF cluster will percolate can be
well understand from the following quantity, $\chi_{\infty}$ which is
defined as \cite{Stauffer},
\begin{equation}
\chi_{\infty}=\frac{Sites\hspace{1mm}in\hspace{1mm} a\hspace{1mm}
  spanning\hspace{1mm} cluster}{Total \hspace{1mm}number\hspace{1mm}
  of\hspace{1mm}occupied \hspace{1mm}sites}
\label{chi}
\end{equation}
\\ \indent Here we briefly outline the idea used in the HK algorithm
for the sake of completeness. It is based on the special application
of the union-find algorithm where it aims to assign a label to each
cluster. For that, we represent our 2D lattice as a matrix and label
all occupied sites initially with -1 and unoccupied sites with 0. Our
cluster index starts with 1. Corresponding to each occupied site,
every time we check the neighbours at the top and left corner of the
current site. If both sites are empty, we label a new cluster number
that has not used so far. Else, if the site has one occupied
neighbour, then we assign same cluster number to the current site. If
both neighbouring sites are occupied, then we set the smallest number
cluster label of the occupied neighbours to use as the label for the
current site. To link two clusters, we create a union between both
labels and set the site as the lowest of the two labels. When we burn
the lattice for a second time, we collect the unions and update the
lattice.
\begin{figure*}[!t]
  \centerline{
    \hfill \psfig{file=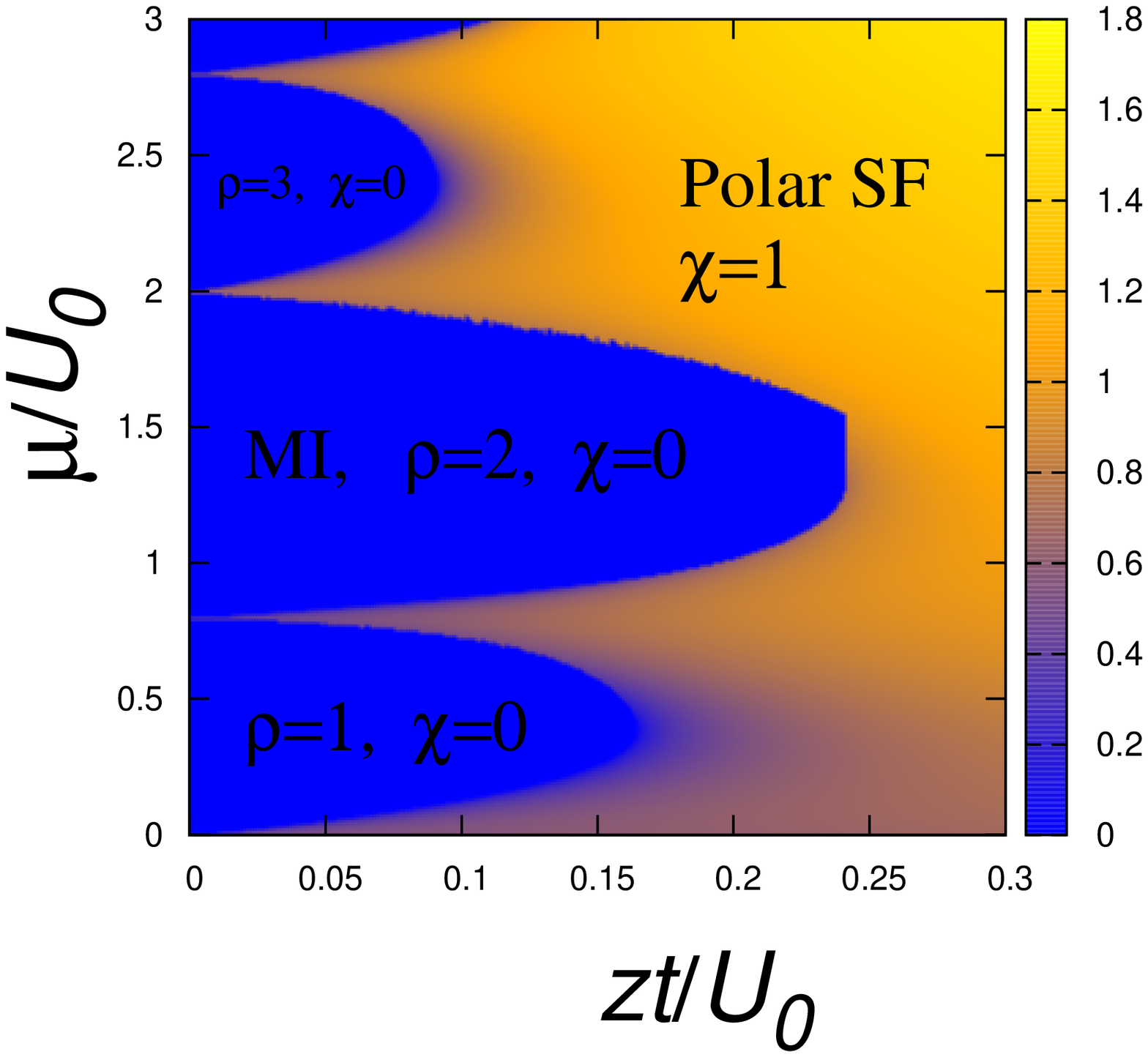,width=0.29\textwidth}
    \hfill
    \hfill \psfig{file=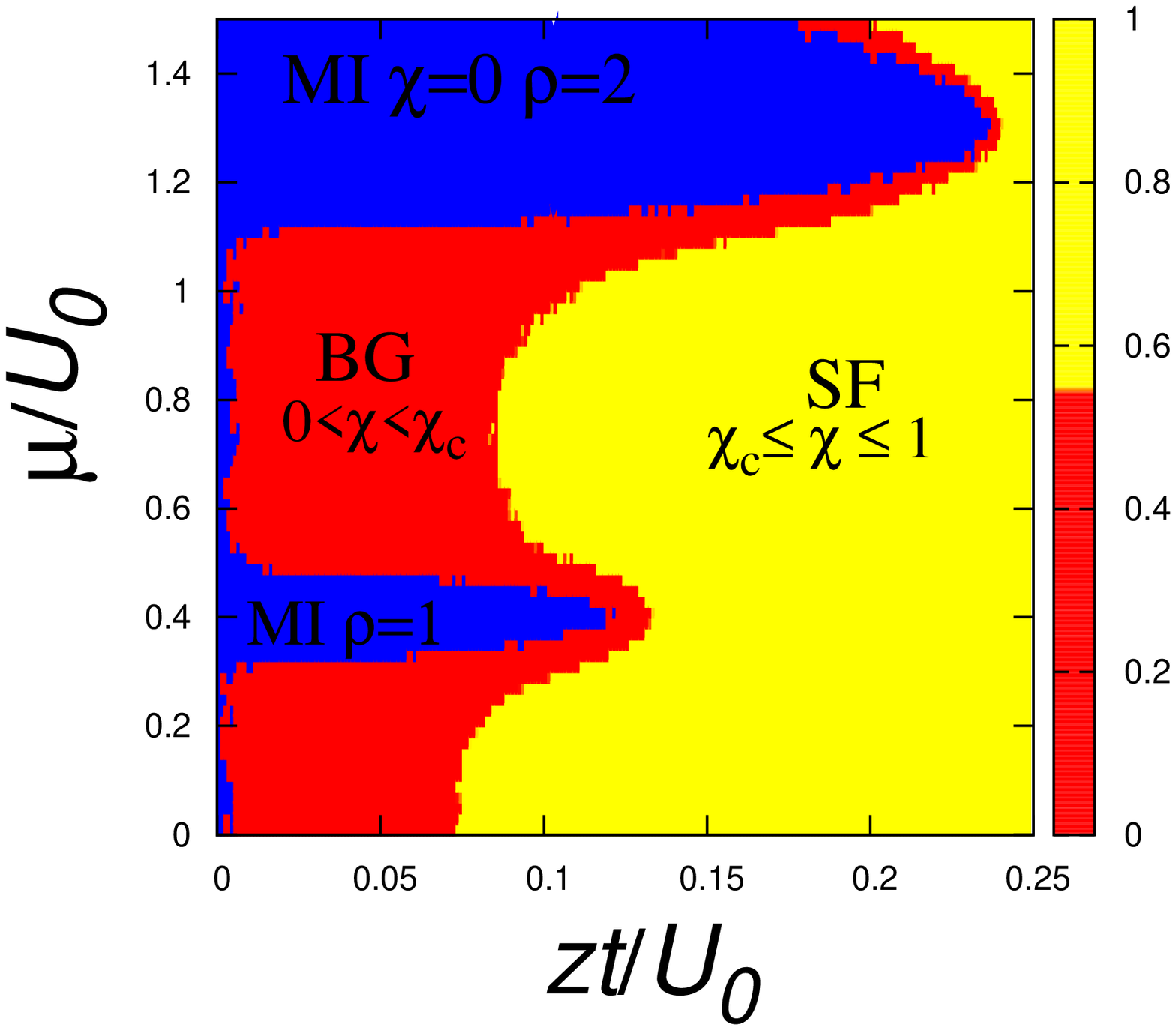,width=0.3\textwidth}
    \hfill
    \hfill \psfig{file=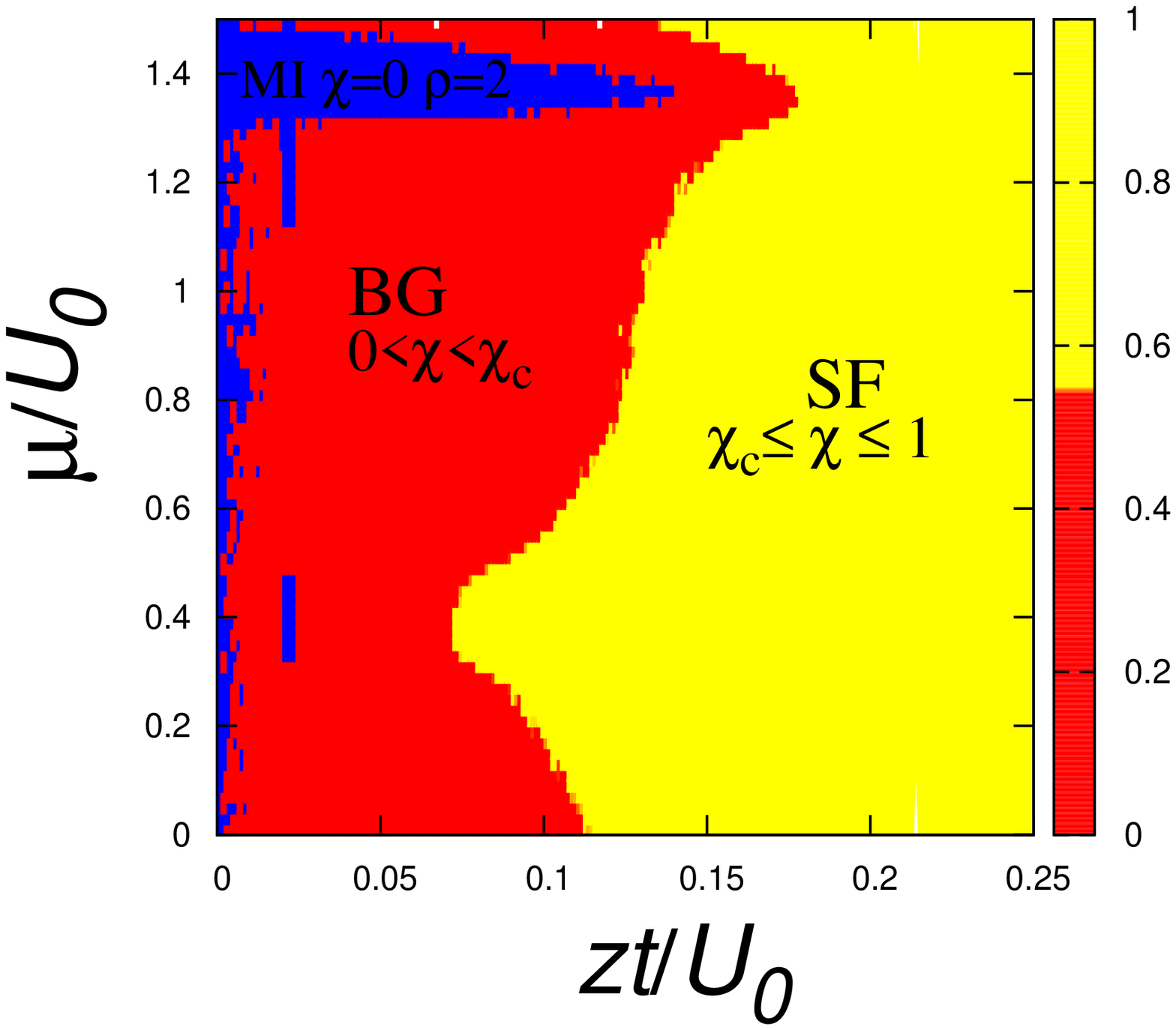,width=0.29\textwidth}
    \hfill}
   \centerline{$(a)$\hspace{5cm}  $(b)$ \hspace{5cm} $(c)$} 
  \caption{(Color online) Phase diagram of spinor ultracold atoms in
    $t-\mu$ plane at $U_{2}/U_{0}$=+0.1 (antiferromagnetic). In the
    pure case ($\Delta/U_{0}$=0.0), the odd MI lobes start to shrink
    while the even MI lobes expanded and the colour variation
    corresponds to the magnitude of the SF order parameter $\psi$
    shown in the legend of (a). Phase diagram based on the information
    obtain from $\chi$ for the disordered cases, namely $\Delta/U_{0}$
    =0.3 is shown in (b) and $\Delta/U_{0}$=0.5 in (c).}
\label{phasepolar}
\end{figure*}
  \\ \indent In Fig.[\ref{finitesize}(a)] we study
$\chi_{\infty}$ as a function of $\chi$ at a disorder strength
$\Delta/U_{0}$=0.5 and $\mu/U_{0}$=0.4 in the antiferromagnetic
case. It is found that $\chi_{\infty}$ is zero till
$\chi$=$\chi_{c}$=0.581 for a lattice size $L\times L$ =$128\times128$
which is close to the critical threshold value for the occupation
probability $p_{c}$=0.592 for random site percolation problem in an
infinite 2D square lattice \cite{Stauffer} and shows finite value
above $\chi>\chi_{c}$. These similarity is only coincidental as the
random site percolation model is a classical problem where sites are
randomly occupied, while we have a system of quantum particles and the
driving parameters are systematically varied at a given value of
random disorder.  \\ \indent Another useful quantity in this regards,
is the mean cluster size, $M_{cs}$ defined as \cite{Stauffer}-
\begin{equation}
M_{cs}(\chi)=\frac{\sum\nolimits_{p}^{\infty}p^{2}s_{p}(\chi)}{\sum\nolimits_{p}^{\infty}ps_{p}(\chi)}
\label{mcs}
\end{equation}
where $ps_{p}$ is the number of occupied sites belonging to a $p$-$th$
cluster and the spanning clusters are excluded from the sum. Since the
spanning clusters are excluded from the sum, the mean cluster size
continues to increase with $\chi$ till the appearance of a spanning
cluster for the first time and starts to decay immediately after
that. The variation of $M_{cs}$ with $\chi$ at a disorder strength
$\Delta/U_{0}$=0.5 for $\mu/U_{0}$=0.4 [see Fig.\ref{finitesize}(b)] shows
that it reaches its peak value just below the $\chi_{c}$ and falls off
after $\chi_{c}$ which confirms that the system is in the BG phase
till $\chi_{c}$, and beyond that it goes toward the SF phase.  \\ \indent
To deal with the finite size effects and also to determine the
percolation transition, we resort to the finite size scaling. For a
given system size $L$, we shall now study the percolation probability
$P_{perc}$ which is the probability of having a percolating cluster
with the system parameter $\eta$=$zt/U_{0}$. Fig.[\ref{finitesize}(c)]
shows the variation of $P_{perc}$ with different lattice sizes $L$ at
a particular disorder strength, $\Delta/U_{0}$=0.5 and $\mu/U_{0}$=0.4
in the antiferromagnetic case. We averaged our results over 1000
($L$=10) and 100 ($L$=30, 50) different realizations of disorder. The
$P_{perc}$ is assumed to follow a scaling law near the critical
tunneling strength, $\eta_{c}$ \cite{Stauffer} and is described by,
\begin{equation}
P_{perc}(L,\eta)=\tilde{p}(L^{\frac{1}{\nu}}(\eta-\eta_{c}))
\end{equation}
where $P_{perc}$ for different $L$ values intersect approximately at
one critical point given by $\eta_{c}$=$zt_{c}/U_{0}$ and $\tilde{p}$
is the scaling function which approaches zero in $(\eta-\eta_{c})<<0$
and unity for $(\eta-\eta_{c})>>0$, $\nu$ is the critical exponent
which is equal to 1.33 for conventional random site percolation
problem in two dimension. The finite size scaling plot, where all data
corresponding to different system sizes collapse on a single curve, is
shown in Fig.[\ref{finitesize}(d)] where the corresponding critical
tunneling strength is given by $\eta_{c}$=$zt_{c}/U_{0}$=0.073. For
the sake of brevity we have not included the discussion corresponding
to the ferromagnetic case.
\begin{figure*}[!t]
  \centerline{
    \hfill \psfig{file=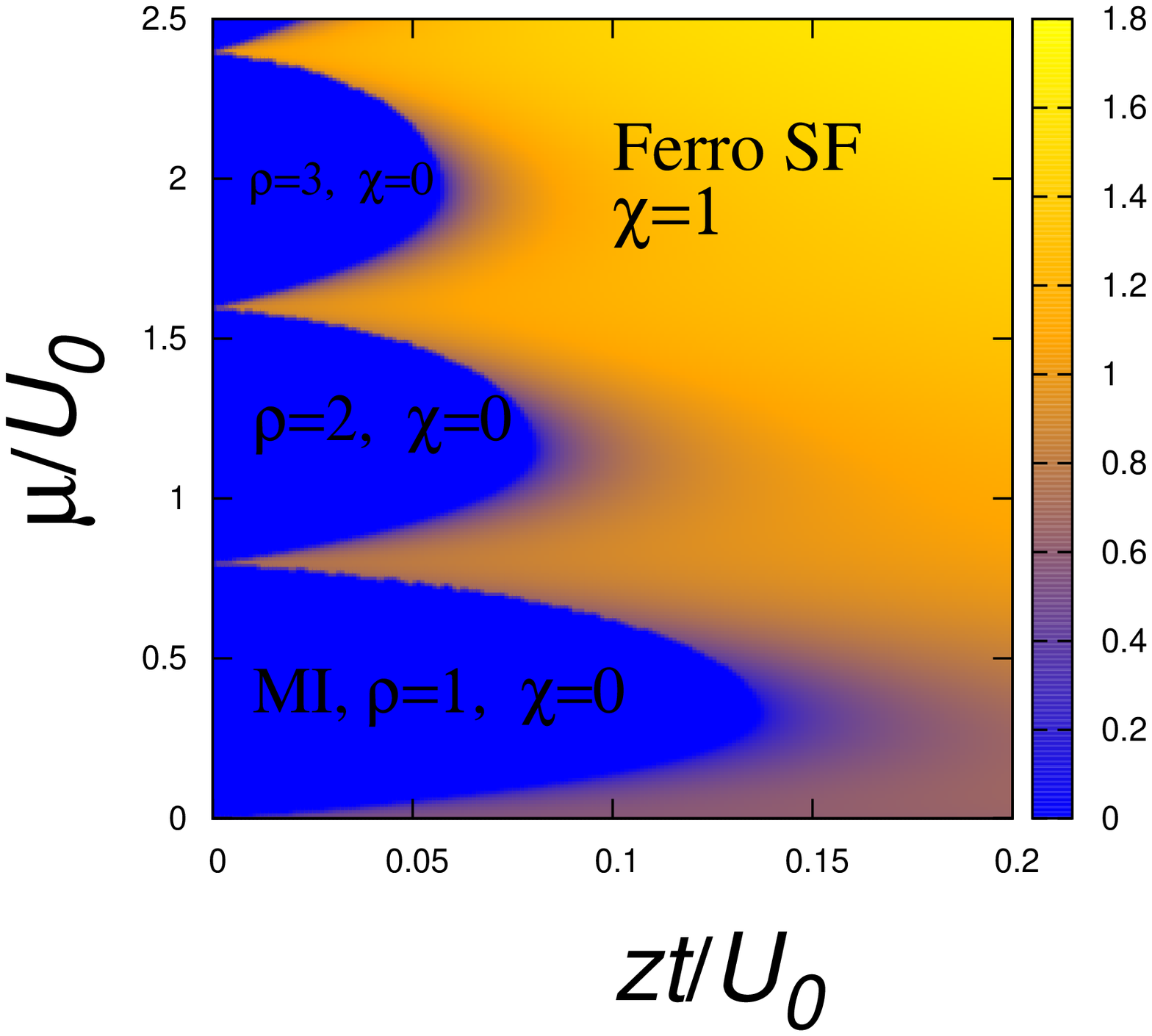,width=0.3\textwidth}
    \hfill
    \hfill \psfig{file=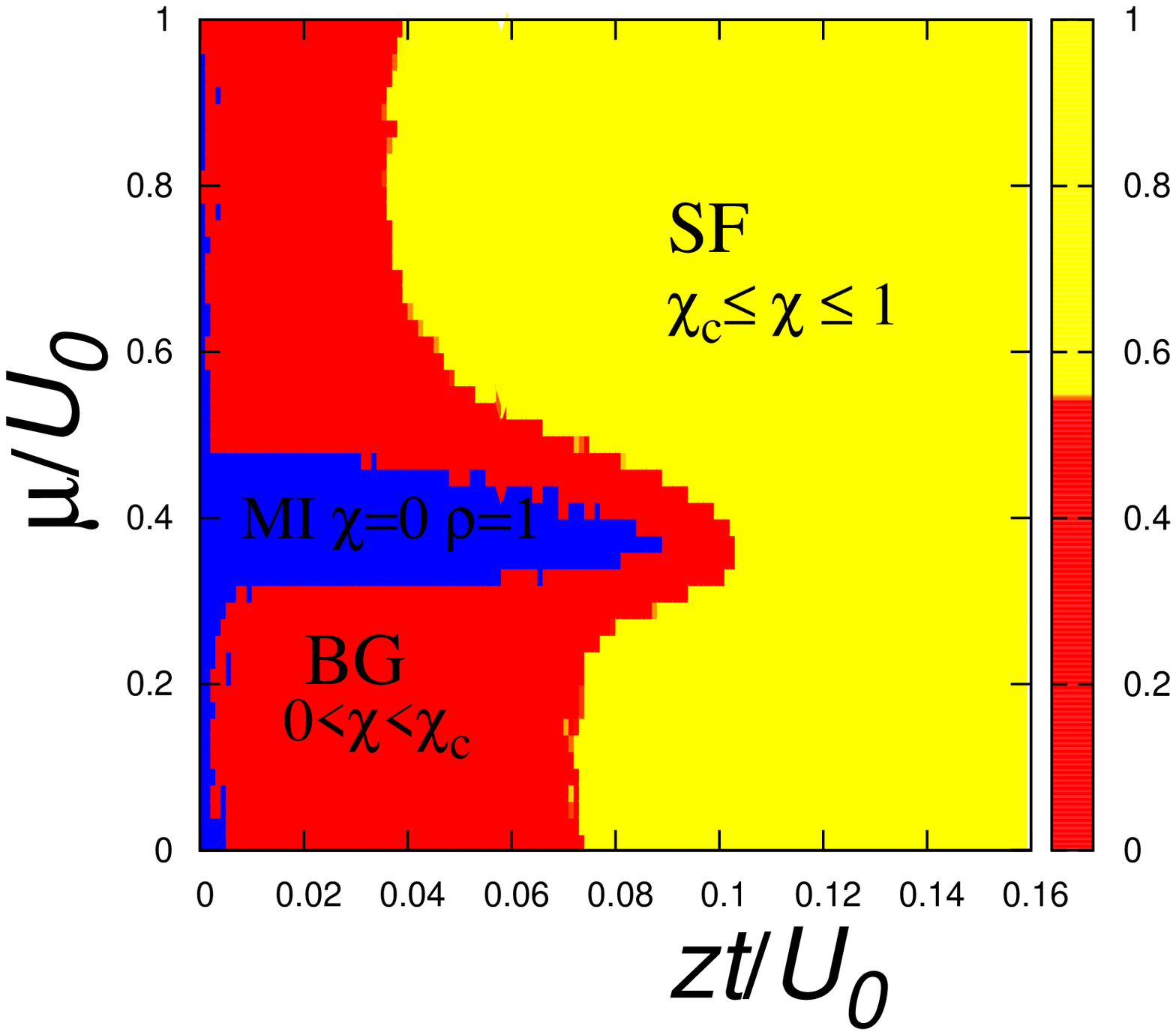,width=0.3\textwidth}
    \hfill
    \hfill \psfig{file=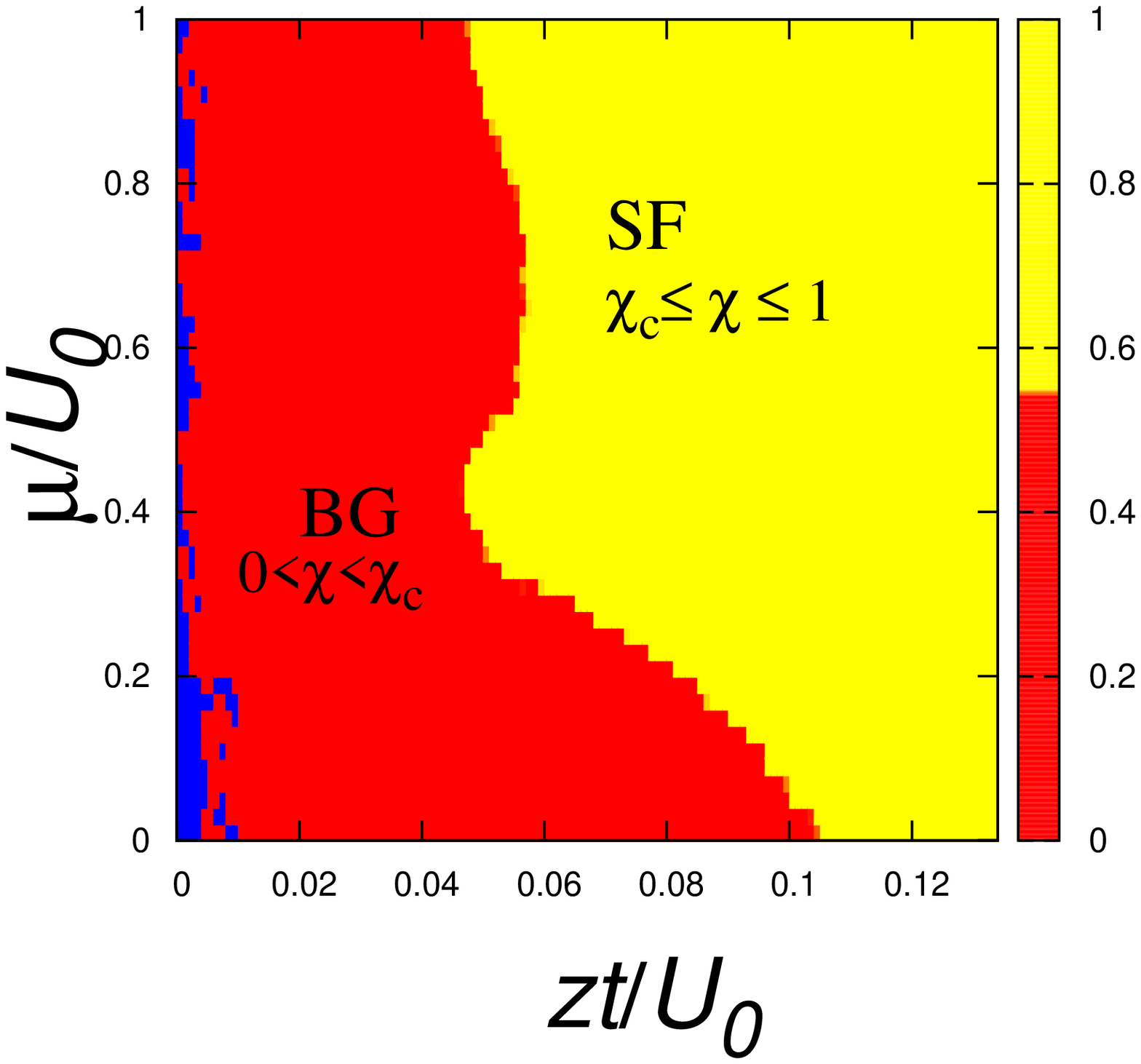,width=0.3\textwidth}
    \hfill}
    \centerline{$(a)$\hspace{5cm}  $(b)$ \hspace{5cm} $(c)$} 
  \caption{(Color online) Phase diagram of spinor ultracold atoms in
    $t-\mu$ plane at $U_{2}/U_{0}$=-0.2 (ferromagnetic). In the pure
    case ($\Delta/U_{0}$=0.0), the phase diagram is similar to that of
    scalar Bose gas (a). Phase diagram based on the information obtain
    from $\chi$ for the disordered cases, namely $\Delta/U_{0}$ =0.3
    is shown in (b) and $\Delta/U_{0}$=0.5 in (c).}
  \label{phaseferro}
\end{figure*}
\begin{table*}[!t]
\begin{center}
\begin{tabular}{|ccc|c|c|c|c|}
\hline
   &  &  &\underline{$\Delta/U_{0}$=0} &\underline{$\Delta/U_{0}$=0.3} &\underline{$\Delta/U_{0}$=0.5} \\
   &  &  & MI-SF   &MI-BG\hspace{2mm}   BG-SF  & MI-BG \hspace{2mm} BG-SF\\
\hline  
  Interactions\hspace{9.5mm}\vline &MI lobes\hspace{5mm}\vline &$\mu/U_{0}$   &$zt/U_{0}$	&$zt/U_{0}$\hspace{5mm} $zt/U_{0}$  &$zt/U_{0}$\hspace{5mm} $zt/U_{0}$\\
\hline
 AF ($U_{2}/U_{0}$=0.1)\hspace{2.4mm}\vline  &\begin{tabular}{@{}c@{}}{\color{blue}first odd} \\ {\color{red}first even} \end{tabular}\hspace{4.3mm}\vline &\begin{tabular}{@{}c@{}}{\color{blue}0.4} \\ {\color{red}1.4} \end{tabular} &\begin{tabular}{@{}c@{}}{\color{blue}0.16} \\ {\color{red}0.26} \end{tabular} &\begin{tabular}{@{}c@{}}{\color{blue}0.122} \\ {\color{red}0.210} \end{tabular}\hspace{3mm} \begin{tabular}{@{}c@{}}{\color{blue}0.133} \\ {\color{red}0.225} \end{tabular} &\begin{tabular}{@{}c@{}}{\color{blue}No} \\ {\color{red}0.123} \end{tabular}\hspace{3mm} \begin{tabular}{@{}c@{}}{\color{blue}0.073} \\ {\color{red}0.174} \end{tabular}\\
\hline
  F ($U_{2}/U_{0}$=-0.2)\hspace{3.8mm}\vline  & first\hspace{11.8mm}\vline &0.4  &0.133  &0.085\hspace{4mm}0.103     &No \hspace{6mm}0.047\\
\hline 
\end{tabular}
\caption{The transition points for the MI-BG and BG-SF phases obtained
  from $\chi$ and $\chi_{\infty}$ for both antiferromagnetic (AF) and
  ferromagnetic (F) interactions are presented in this table. The
  corresponding parameter values are mentioned in the table. }
\label{table1}
\end{center}
\end{table*}
\\ \indent We are now in a position to distinguish the three different
types of phases and able to ascertain the critical tunneling strength,
$\eta_{c}$=$zt/U_{0}$ corresponding to the MI-BG and the BG-SF phase
transitions. As we have already pointed out earlier that the MI regime
corresponds to $\chi$=0, while the BG regime corresponds to
$0<\chi<\chi_{c}$ till $\chi_{\infty}$=0 and the SF regime is given
by, $\chi_{c}\le \chi$ $\le 1$ when $\chi_{\infty}\ne$0. For that
purpose we shall tune $zt/U_{0}$ in a controlled manner to closely
monitor the value of $\chi$ and $\chi_{\infty}$ for a given lattice
size namely, $L$=30. We find that in the antiferromagnetic case (at
$\mu/U_{0}$=0.4), $\chi$ is zero till $zt/U_{0}$=0.122 at
$\Delta/U_{0}$=0.3, while it is non zero at higher disorder
concentration $\Delta/U_{0}$=0.5. Thus at $\mu/U_{0}$ =0.4, the onset
for the MI-BG transition occurs at $zt/U_{0}$=0.122 and the BG phase
extends upto $zt/U_{0}$=0.133, where $\chi_{c}$=0.524 with
$\chi_{\infty}$=0. Hence the BG-SF transition takes place at
$zt/U_{0}$=0.133 corresponding to $\Delta/U_{0}$=0.3. For
$\Delta/U_{0}$=0.5, since there is no MI phase, so the BG-SF phase
transition occur at $zt/U_{0}$=0.073 with $\chi_{c}$=0.549.
\\ \\\textbf{D. Phase diagram} \\ \indent In this section we will now
present the phase diagram of spinor ultracold atoms in $t-\mu$ plane
for both the pure and disorder cases based upon the information
obtained from $\psi_{i}$, $\chi$ and $\chi_{\infty}$. In the pure
case, the phase diagram can be easily obtained from the information of
the SF order parameter, $\psi_{i}$, while in the disordered case, we
determine our phase diagram depending upon the range of $\chi$ as
discussed earlier corresponding to both the antiferromagnetic and
ferromagnetic cases.  \\\\ \indent (i) Antiferromagnetic case: In the
disorder free case, the phase diagram of spinor ultracold atoms in
presence of antiferromagnetic interaction with $U_{2}/U_{0}$=+0.1 is
shown in Fig.[\ref{phasepolar}(a)]. It shows that the MI lobes with
even occupation densities are somewhat enhanced as compared to the
corresponding lobes with odd occupation densities \cite{Pai,
  Tsuchiya}. The MI-SF phase transition for the even MI lobes is first
order in nature, while it is second order for the odd MI lobes, as
obtained from the variation of the SF order parameter and
compressibility which was outlined in the preceding section. The MI
phase with even number of bosons per site where each pair of bosons
favour the formation of singlet parts and likely are to the more
stable and keep themselves isolated from nearest neighbour
tunneling. Whereas the MI phase, with odd number of bosons per site,
although each pair of bosons can make singlet states, the remaining
one is free to tunnel to nearest neighbour sites rendering the odd MI
lobes unstable and smaller in width as compared to the even MI
lobes. \\ \indent In presence of disorder, we now try to obtain our
phase diagram entirely based on the information obtained from
$\chi$. In the previous section, since we distinguished between the
three different kind of phases depending upon $\chi$ and
$\chi_{\infty}$. Thus for obtaining the phase diagram, we set $\chi$=0
as the indicator for the MI phase, $0<\chi<\chi_{c}$ (=0.524) for the
BG phase and $0.524\le\chi \le 1$ for the SF phase at
$\Delta/U_{0}$=0.3. Similarly for $\Delta/U_{0}$=0.5, where $\chi_{c}$
is equal to 0.549. The phase diagrams for both the disorder strengths
are shown in Figs.[\ref{phasepolar}(b) and \ref{phasepolar}(c)]. At
$\Delta/U_{0}$=0.3, the intervening BG region for the odd MI lobes is
more noticeable as compared to the even MI lobes
[see Fig.\ref{phasepolar}(b)]. At $\Delta/U_{0}$=0.5, the BG phase
completely destroys the first odd MI lobe, but the even MI lobe still
exists as expected from the earlier discussion
[see Fig.\ref{phasepolar}(c)]. \\\\ \indent (ii) Ferromagnetic case: In
Fig.[\ref{phaseferro}(a)], we plot our numerically obtained phase
diagram corresponding to the pure case with a ferromagnetic
interaction namely, $U_{2}/U_{0}$=-0.2. The phases bear similar
characteristic features as that of scalar or spinless Bose gas. All
the three SF spinor components are non zero in the SF phase which
smoothly vary across the MI to SF phases and hence show a second order
phase transition. In the disordered case, as earlier, we determine the
phase diagram by invoking the critical value of $\chi_{c}$=0.5034 at
$\Delta/U_{0}$=0.3 and $\chi_{c}$=0.557 at $\Delta/U_{0}$=0.5
corresponding to the onset of the SF phase. The phase diagram
corresponding to both the disorder values are shown in
Figs.[\ref{phaseferro}(b) and \ref{phaseferro}(c)]. For $\Delta/U_{0}$=0.5, the MI
lobes disappear since the disorder strength is higher than the
critical disorder strength and the system is left only with the BG and
the SF phases.
\section{Conclusions}
We have studied the effect of onsite disorder in a two dimensional
SBHM for both the antiferromagnetic and ferromagnetic
interactions. The appearance of the BG phase is observed via the
average SF order parameter $\bar\psi$ and compressibility
$\bar\kappa$. The $\bar\psi_{\sigma}$ are zero while
$\bar\kappa_{\sigma}$ are zero or constant in MI phase and gradually
increases with disorder yielding signature for a BG phase. In the
antiferromagnetic case, the MI- SF phase transition for odd MI lobes
is always second order while for even MI lobes, it shows second order
phase transition with increasing disorder strength. We also find that
the MI phase exists till the disorder strength is below the critical
value, while it vanishes above the critical value. \\ \indent Further
the three different types of phases, namely the MI, BG and SF phases
are identified based on the concept of the SF percolating clusters. We
define the MI phase where no SF cluster exists and the SF phase with
at least one SF percolating cluster exists. The BG region corresponds
in between the MI and SF phase that is the SF cluster exists but does
not percolate. We locate the transition point for the MI-BG and the
BG-SF phase by calculating the probability of having a percolating
cluster $\chi_{\infty}$ ($P_{perc}$). A percolation analysis of SF
percolating cluster is studied using HK algorithm with different
system sizes $L$ as a function of tunneling strength,
$\eta$=$zt/U_{0}$ which obey finite size scaling law. \\ \indent
Finally, we obtain our phase diagram based on $\chi$ for the
antiferromagnetic and ferromagnetic cases. For that purpose, we
determine the critical value of $\chi$ where a SF clusters percolate
first time throughout the lattice and set the respective ranges for
$\chi$ for three different phases. Thus a reliable enumeration of different phases can be obtained at desired parameter values of the SBHM Hamiltonian. Since these parameters can be precisely controlled in experiments using the properties of the laser fields which superimpose to form optical lattices, our study may have significant impact on the experimental scenario to determine these phases \cite{Jaksch}.\\ \indent The transition points
for the MI-BG and BG-SF phases obtained from $\chi$ and
$\chi_{\infty}$ are presented in the table \ref{table1} for both
antiferromagnetic (AF) and ferromagnetic (F) interactions in pure and
disordered cases. \\ \indent Now we shall include some comments
concerning the drawbacks in the mean field theory. Since our work is
mainly a focus on the effects of disorder in SBHM, thus the single
site mean field theory always is not able to handle the site
inhomogeneity of the system properly and it works well in higher
dimension $d\ge 2$. Although we take into the effect the site
inhomogeneity by defining $\chi$, but ignoring quadratic fluctuations
introduces a tiny error in the calculations. For example, the MI phase
will vanish above the critical disorder strength, our results are in
good agreement against this claim within a 3\% error.

\section{Acknowledgments}
We thank R. V. Pai for useful discussions. SNN likes to thank
A. Barman for valuable discussions. SB thanks CSIR, India for
financial support under the grants F no: 03(1213)/12/EMR-II.

\end{document}